\documentclass[article]{elsarticle}

\usepackage{hyperref}
\usepackage{lineno}
\modulolinenumbers[5]

\journal{Computer Physics Communications}

\begin{document}

\begin{frontmatter}

\title{REST-for-Physics, a ROOT-based framework for event oriented data analysis and combined Monte Carlo response.\\ \includegraphics[height=25mm]{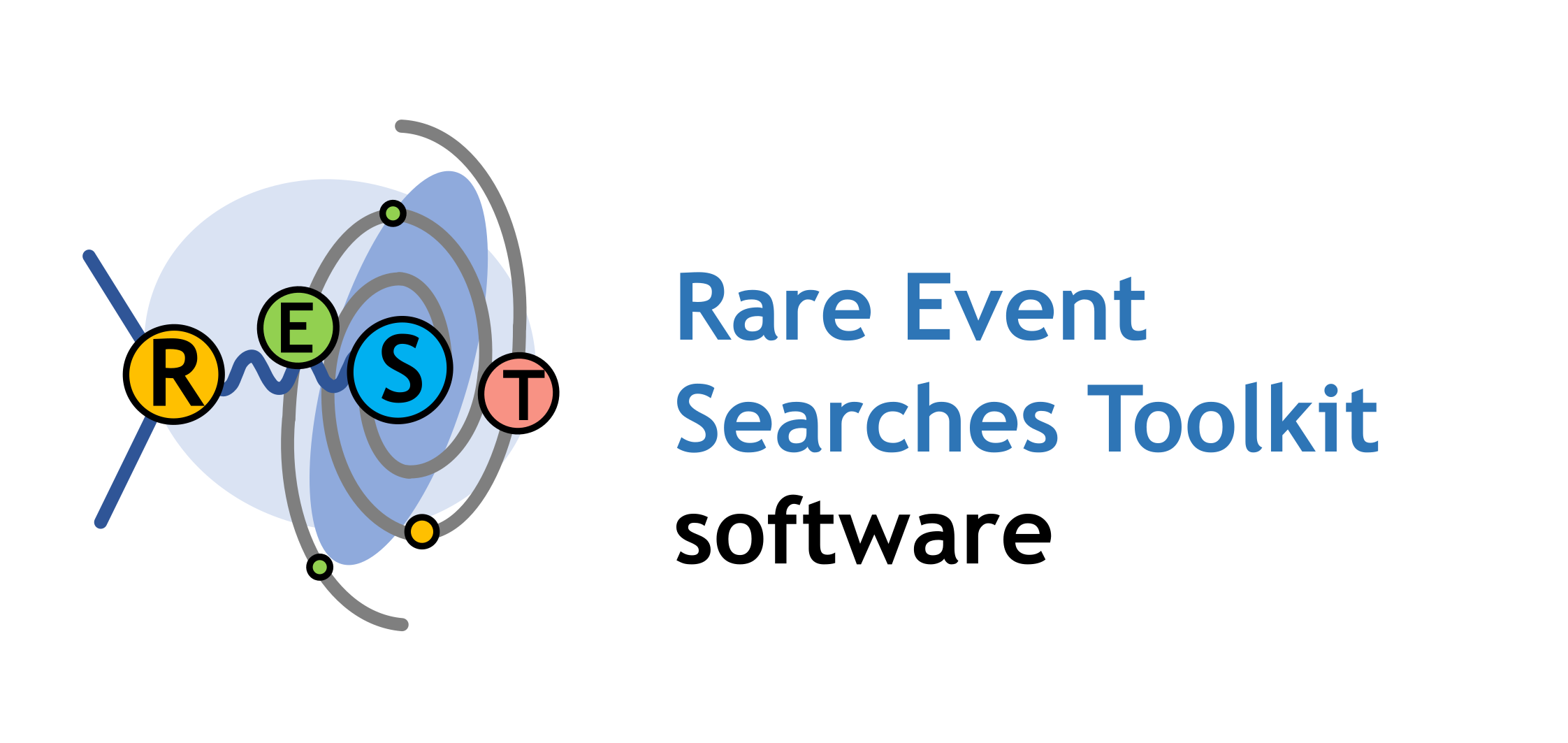}}

\author{Konrad~Altenm\"uller}
\author{Susana~Cebri\'an}
\author{Theopisti~Dafni}
\author{David~D\'iez-Ib\'añez}
\author{Javier~Gal\'an\corref{mycorrespondingauthor}}
\ead{javier.galan@unizar.es}
\author{Javier~Galindo}
\author{Juan~Antonio~Garc\'ia}
\author{Igor~G.~Irastorza}
\author{Gloria~Luz\'on}
\author{Cristina~Margalejo}
\author{Hector~Mirallas}
\author{Luis~Obis}
\author{Oscar~P\'erez}

\address{Center for Astroparticles and High Energy Physics (CAPA), Universidad de Zaragoza, 50009 Zaragoza, Spain}

\author{Ke~Han}
\author{Kaixiang~Ni\corref{mycorrespondingauthor}}
\ead{bur\_ning@sjtu.edu.cn}
\address{INPAC; Shanghai Laboratory for Particle Physics and Cosmology; Key Laboratory for Particle Astrophysics and Cosmology (MOE), School of Physics and Astronomy, Shanghai Jiao Tong University, Shanghai 200240, China}


\author{Yann~Bedfer}
\author{Barbara~Biasuzzi}
\author{Esther~Ferrer-Ribas}
\author{Damien~Neyret}
\author{Thomas~Papaevangelou}
\address{IRFU, CEA, Universit\'e Paris-Saclay, F-91191 Gif-sur-Yvette, France}

\author{Cristian~Cogollos}
\author{Eduardo~Picatoste}
\address{Institut de Ci\`encies del Cosmos, Universitat de Barcelona, Barcelona, Spain}
\address{Departament de F\'isica Qu\`antica i Astrof\'isica, Universitat de Barcelona, Barcelona, Spain}


\cortext[mycorrespondingauthor]{Corresponding author}





\begin{abstract}

The REST-for-Physics (Rare Event Searches Toolkit for Physics) framework is a ROOT-based solution providing the means to process and analyze experimental or Monte Carlo event data. Special care has been taken to the traceability of the code and the validation of the results produced within the framework, together with the connectivity between code and stored data, registered through specific version metadata members.

The framework development was originally motivated to cover the needs of Rare Event Searches experiments (experiments looking for phenomena having extremely low occurrence probability, like dark matter or neutrino interactions or rare nuclear decays). The framework components naturally implement tools to address the challenges in these kinds of experiments. The integration of a detector physics response, the implementation of signal processing routines, or topological algorithms for physical event identification are some examples. Despite this specialization, the framework was conceived thinking in scalability. Other event-oriented applications could benefit from the data processing routines and/or metadata description implemented in REST, being the generic framework tools completely decoupled from dedicated libraries.

REST-for-Physics is a consolidated piece of software already serving the needs of different physics experiments - using  gaseous Time Projection Chambers (TPCs) as detection technology - for detector data analysis and characterization, as well as generic R\&D. Even though REST has been exploited mainly with gaseous TPCs, the code could be easily applied or adapted to other detector technologies. We present in this work an overview of REST-for-Physics, providing a broad perspective to the infrastructure and organization of the project as a whole. The framework and its different components will be described in the text.


\end{abstract}


\end{frontmatter}


\section{Introduction}
\label{sec:intro}

REST-for-Physics\footnote{Along the text we may refer to REST-for-Physics as simply REST. The REST-for-Physics naming is preferred to avoid naming conflict with other unrelated but popular software packages.} (Rare Event Searches Toolkit for Physics) is a collaborative software effort providing a common framework and tools for acquisition, simulation, generic data analysis, and detector response in experimental particle physics. An ambitious feature of REST-for-Physics is its capability to analyze and compare both Monte Carlo and experimental data using the same \emph{event processing} routines upon a unified \emph{event data} - \emph{metadata} architecture. 
The framework was born to bring together different software requirements related to gaseous Time Projection Chambers (TPCs) in the context of Rare Event Searches, and to unify and coordinate various independent developments in a common modular infrastructure with potential for scalability and reusability. Special care has been taken to ensure the traceability and reproducibility of the results obtained after the data processing, linking the code version with the metadata version stored on disk, and protecting such relation. Any user local changes to the code are identified at compilation time. This is used to guarantee that the executed version and the results written to disk correspond to an unmodified official public release. This fact is extremely relevant when planning to register official experimental data and preserve it for historical reasons, such as covering the data management plan of scientific collaborations, including the release  of data to be publicly exploited outside the collaboration domain. The code updates are periodically published in the Zenodo citations system, where a reference to the latest official release is found\,\cite{javier_galan_2021_5092550}.

REST-for-Physics is the result of several years of experience on detector physics and research, motivated originally to cover the software needs of the T-REX project for neutrino and dark matter searches~\cite{Irastorza:2015dcb,Irastorza:2015geo}. The REST-for-Physics code has benefited from several academic works, as it becomes apparent in several PhD thesis publications\,\cite{IguazThesis,tomas2013development,SeguiThesis,HerreraThesis,GraciaThesis, GarciaPascualThesis, RuizThesis} that have contributed to shape and define the final project that is described in this manuscript.
This project has contributed to the development of different but interconnected research activities in a coherent way, unifying common tools that are regularly used today not only in research but also at all the academic levels, from undergraduate to master students. 


Different experimental projects have seen REST-for-Physics growing from its preliminary stages to the mature project we present in this work. REST-for-Physics has been evolving within, and it is being used to produce results at, CAST~\cite{Anastassopoulos:2017ftl}, TREX-DM~\cite{trexdm_bckmodel}, PandaX-III~\cite{pandaxiii_cdr,Lin:2018mpd,Galan:2019ake}, and IAXO~\cite{Armengaud:2019uso}. Those projects have benefited from the consolidation of REST as a common tool widely used among collaborators to process, register and analyze detector data. The use and development of REST in other experiments is encouraged in a community effort to maintain appropriate tools for related tasks. In addition to sharing the know-how and experience in our physics domain, the motivation to release a public common framework resides in providing the possibility to distribute the experimental data following a unique format readable with REST-for-Physics, or any other ROOT I/O compatible code, in a future open-data program of the experiments. The code is open-source and it is distributed under a GNU public license at \emph{GitHub}\,\cite{REST_Git}.

The aim of this document is to give the reader a broad perspective of the purpose of the software project, its organization and contents, and the basic instruments that shape the whole infrastructure, giving an idea of its scalability potential, and in addition, showing the code validation strategy and continuous integration philosophy. For further reference, detailed information is provided, including an API class documentation for developers\,\cite{REST_API}  synchronized daily with the latest development version, and a comprehensive guide for first time users\,\cite{REST_user_guide}. An additional communication channel is available in the form of a public forum\,\cite{REST_forum} to encourage discussion about topics related to our field, help others on their first steps using REST and/or integrate their first routines inside the framework, and discuss about new or existing feature upgrades.

The REST-for-Physics potential resides on its capability to be used with Monte Carlo or experimental data, or a combination of both in an event processing chain. REST is used for modeling, simulation and/or detector response, but not exclusively. The aim of this document is not to provide a detailed description of particular calculations, but to provide an overall description of the existing tools that are used frequently on such duty. This document is distributed as follows. In section\,\ref{sec:conception} we provide the framework philosophy from a conceptual perspective, the contextualization of the environment where REST was born and the scope of the project itself. In section\,\ref{sec:framework} we give a broad description of the main framework infrastructure, the basic concepts and/or elements that shape its behavior, common analysis and visualization tools, and job management. In section\,\ref{sec:libraries} we introduce the most common libraries that implement dedicated algorithms for specific tasks in the aforementioned duties.
\section{REST conceptual design and scope}
\label{sec:conception}

REST-for-Physics defines common data structures for event-based data processing. As it will be seen later in the general description (in section\,\ref{sec:framework}), this entails a prototyping of the event data holder, the processes that transform or operate those data holders, and the description of the metadata information giving a meaning to the data being processed: initial data taking conditions, input processing parameters, output results written to disk in the form of metadata, etc.
The prototypes of event data, processes and metadata are complemented  with basic analysis tools that are frequently used on event-based data analysis. Another important structure, named tree, is used to gather relevant event information during the data processing. This analysis summary tree contains a set of variables defined during the event data processing to be used in subsequent, higher level analysis.

REST-for-Physics defines a framework, or code development space, that centralizes event processing and analysis routines. These routines contributed by the same experts that work on the analysis of the data. The REST community keeps a strong link between  algorithm design and the framework design, since there is an implicit connection between the algorithm development, analysis interpretation, and framework design requirements. REST provides already existing processes that can be used directly to define a given event processing task. REST has been designed to provide the means to be extended with new processes, metadata or event data types.

The development of REST emerges in a strong academical environment. In such context, REST intents to provide the means for academic works to materialize in the form of a piece of code that can be re-used within an already consolidated software infrastructure. A major goal for REST is to make it more accessible to non-computing experts that have a high level for algorithm coding abstraction and comprehension of the physics context.

REST-for-Physics does not replace nor does it compete with other dedicated simulation packages which provide high accuracy physics description on dedicated problems; it seeks to integrate those packages, such as Geant4~\cite{Agostinelli:2002hh} or Garfield++~\cite{Garfield}, and exploit them inside the framework as needed on the processing of the event data. In addition, REST-for-Physics includes dedicated libraries (described in section\,\ref{sec:libraries}) that implement specialized algorithms for signal processing or physical track reconstruction. REST has its own algorithms for known mathematical problems, e.g. time signal processing, to have full control over those  and adapt them to our experimental needs, while still linking to consolidated libraries when possible, as it is for example the case for high-precision numbers implementation at the mpfr library\,\cite{10.1145/1236463.1236468} or the use of graph theory methods\,\cite{Applegate:2007:TSP:1374811,concorde}.
\section{General description}
\label{sec:framework}

REST-for-Physics is composed of a set of libraries written in C++ and it is fully integrated with ROOT~\cite{ROOT,Brun:2011Gp,ROOT2011}, i.e. most C++ classes inherit from a ROOT TObject and therefore they can be read, accessed or written using the ROOT I/O interface. The only structural dependence is related to ROOT libraries, while other packages, as Geant4\,\cite{Agostinelli:2002hh} or Garfield++\,\cite{Garfield}, can be optionally integrated and used within the REST-for-Physics framework when generating or processing Monte Carlo data. Since REST-for-Physics is a natural extension of ROOT, the same naming conventions are followed: the Taligent rules. On top of those standard naming conventions, any REST-for-Physics C++ class will always start with the \emph{TRest} prefix. In this paper we will highlight the words when they clearly make a reference to an existing REST-for-Physics C++ class: a class named \emph{TRestEvent} will be written as \emph{event} and a class named \emph{TRestAnalysisTree} will be written as \emph{analysis tree}. Therefore, a highlighted word, within context, expresses a deep connection with the existing C++ classes in the project.

\subsection{The REST-for-Physics framework}
\begin{figure}[]
  \centering
  \raisebox{-0.5\height}{\includegraphics[trim=0 20 0 0, clip,width=0.95\linewidth]{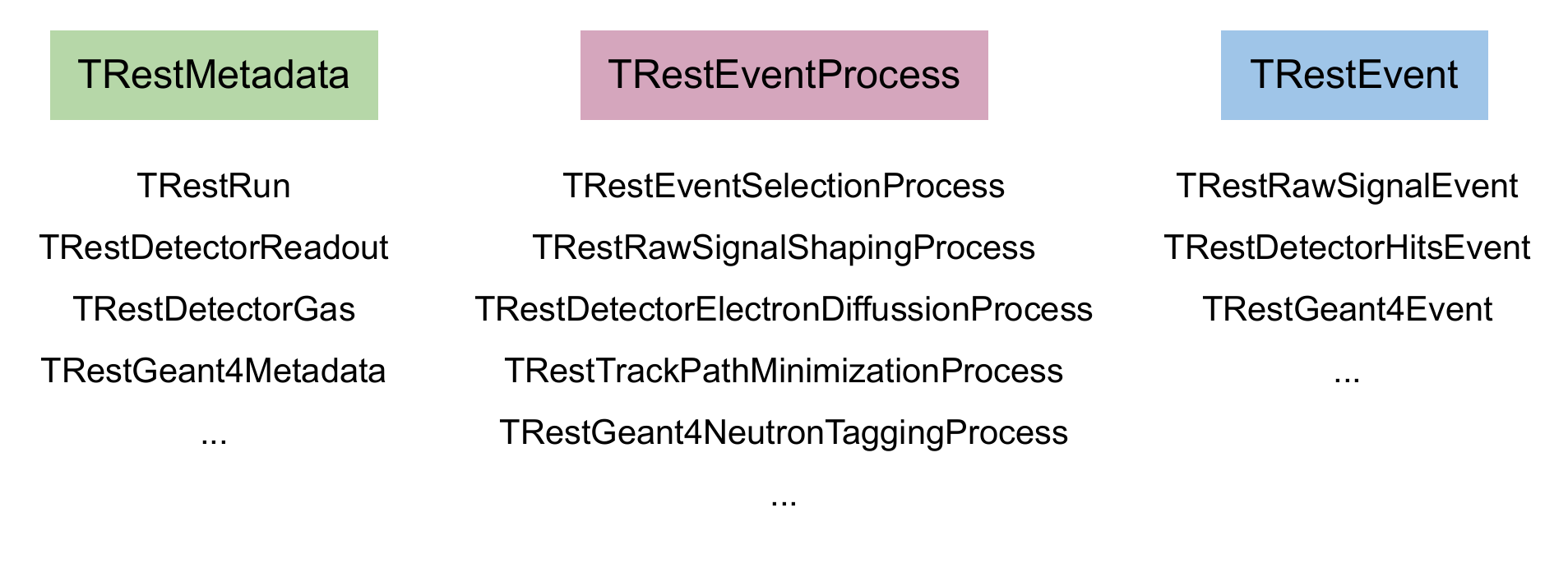}}
	\caption{The base REST-for-Physics framework abstract classes, 
	\emph{metadata}, \emph{event process} and \emph{event} together with few examples of specific class implementations.}  \label{fig:objects}
\end{figure}

Inside the REST-for-Physics ecosystem a core library or framework is found. This core library prototypes and fixes the implementation of most of the REST-for-Physics C++ classes. Those base classes serve to define common methods and data members for specific\footnote{The reader should note that when we refer to specific classes, we refer to classes which inherit, in the strict sense of C++ class inheritance, from the base abstract classes, such as \emph{event}, \emph{metadata} and \emph{event process} classes. In the text, we will highlight the keyword \emph{specific} to refer to those inherited classes in a generic way, e.g. \emph{specific event} will be connected to any \emph{TRestSpecificEvent} inheriting from \emph{TRestEvent}. }
classes (see Figure~\ref{fig:objects}). We shall briefly introduce those basic elements:

\begin{itemize}
    \item The \emph{event} class encapsulates any specific \emph{event} data inside REST. It defines common fields, such as timestamp or event id, and it prototypes common methods used for printing or drawing event information. A particular \emph{specific event} implementation defines a type. Thus it is important to note that in what follows we will distinguish between the \emph{event} data as the explicit contents of a particular \emph{specific event}, and the \emph{event} type as the format, or structure, of the \emph{specific event}. A \emph{specific event} representation is typically a physical quantity that needs to be described in a physical coordinate space or physical time, as it can be the time signals registered by an electronics acquisition system, or the energy deposits distribution produced by a Geant4 simulation.
    
    \item A \emph{metadata} class may be used as a mere information container, storing relevant parameters, such as the description of the simulation conditions in \emph{restG4} (see section\,\ref{sc:geant4lib}). Or it might also adopt the shape of a complex object definition that implements advanced methods, such as the construction of a \emph{detector readout}, or a \emph{magnetic field} volume including interpolation routines. Those advanced \emph{metadata} classes will be found in specialized libraries. Conceptually we understand by \emph{metadata} any information required to give meaning to any \emph{specific event} data. Therefore, any input or output parameters required during the processing or transformation of \emph{event} data, or type, using \emph{event processes}, is also regarded as \emph{metadata}. Any metadata class can be initialized through an Extensible Markup Language (XML) configuration file.
    
    \item The \emph{event process} class defines an input/output \emph{event} protocol allowing to interconnect different \emph{specific event process} implementations into a sequential processing chain. This object (as an instance of a specific class) will be able to perform operations with the input \emph{specific event} transforming its type and/or its contents; the changes being returned in the output \emph{specific event}. The \emph{event process} itself inherits from the \emph{metadata} class, since a process usually requires initialization to define input parameters that control the behavior of the process.
\end{itemize}

Other elemental tools are found inside the main framework, such as string helper methods, fundamental physics constants and units system or other basic mathematical tools useful for the development of any specific \emph{event process}. Any \emph{specific metadata} or \emph{specific event process} class that does not require specialization will likely be hosted inside the framework domain.
 In addition, the framework repository\,\cite{REST_Framework_Git} centralizes other REST-for-Physics components, such as libraries or packages. Those components will be introduced in section\,\ref{sec:libraries}.





\subsection{I/O access and metadata storage}

\begin{figure}[tb]
  \centering
  \raisebox{-0.5\height}{

  \includegraphics[width=0.75\linewidth,trim=0 0 0 0,clip]{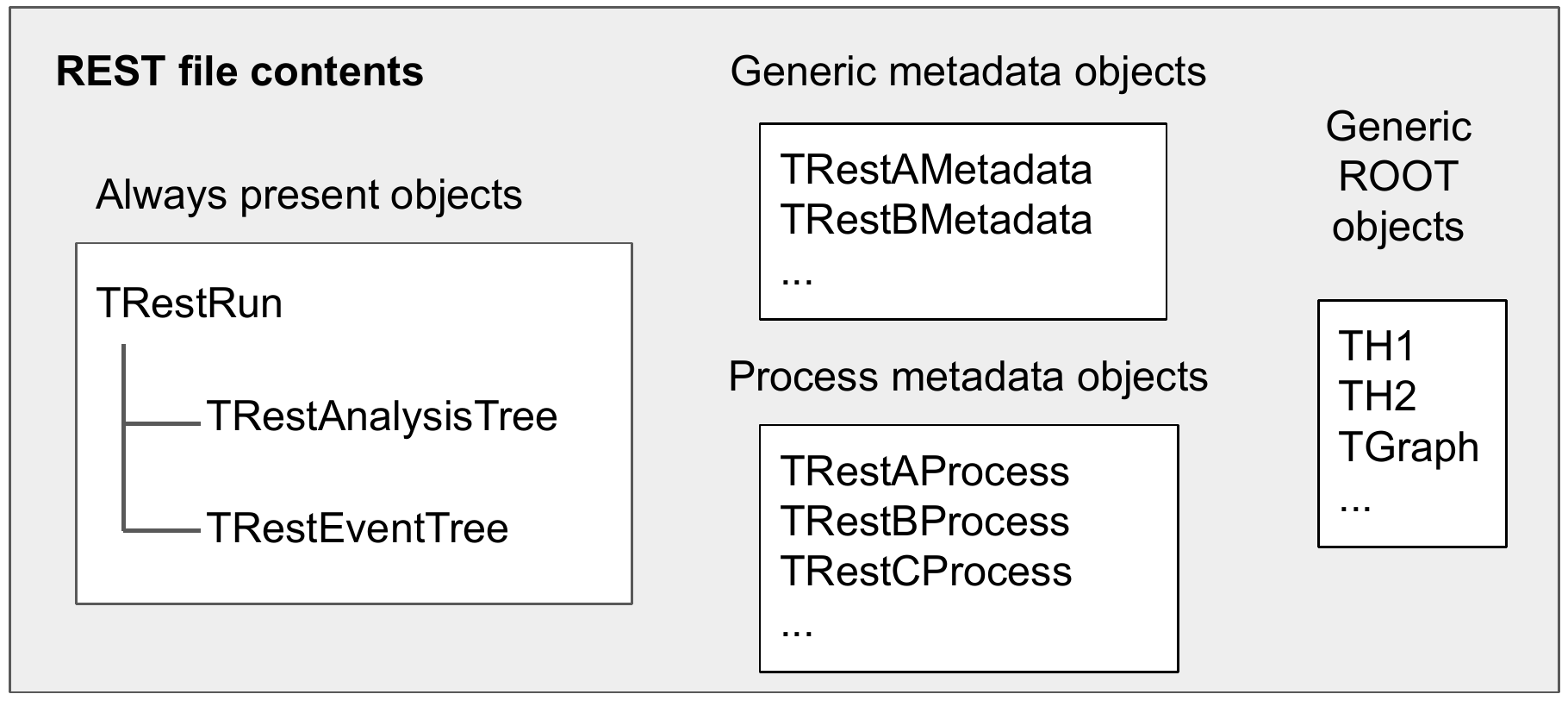}
  }
	\caption{A schematic showing the different components that may be present inside a REST data file. The \emph{analysis tree} and \emph{event tree} objects are independent objects accessed through the \emph{run} interface which ensures coherent access to a particular event entry linked to its corresponding analysis entry.} \label{fig:file_contents}
\end{figure}

REST-for-Physics uses the ROOT I/O interface to write \emph{event} and \emph{metadata} objects to disk. A ROOT file generated with REST may contain any number of \emph{specific event} and \emph{specific metadata} objects, including any \emph{specific event process} (being a \emph{metadata} object itself). Those objects are stored in a unique file, together with the \emph{run} metadata object and the \emph{analysis tree} that are always present in any file that has been processed with REST (see Figure~\ref{fig:file_contents}). The \emph{run} object registers values to identify the data file and the conditions the data were registered, such as the start time, duration, run number, etc, while the \emph{analysis tree} collects per-event information, named observables, at any stage of the data processing. The \emph{run} object takes also an active role when accessing the different objects stored on disk by implementing helper methods to access the data; for example, getting a list of events from the \emph{analysis tree} fulfilling particular conditions or retrieving directly the pointer to a given event id number updating simultaneously its corresponding \emph{analysis tree} entry.

The framework philosophy is to create \emph{specific metadata} classes with dedicated data members to store any information crucial for the final analysis, and/or to fully determine the nature of the data being stored. All the \emph{metadata} inherited objects gain data member reflection support, thus creating a relation between the C++ conceptual class members and the text fields used in the configuration files. Through the implementation of the \emph{metadata} object, a dedicated configuration file format for REST has been designed, based on Extensible Markup Language (XML). This upgrade allows the reading of XML files with additional features, such as system environment variables, complex programming instructions, including \emph{if} conditions or \emph{loops}, evaluation of mathematical expressions or even support for arbitrary physical units conversion inside any parameter. A new extension, \emph{rml}, is assigned to this upgraded XML format.

Using the \emph{metadata} philosophy, a unique relation between the configuration files and the C++ objects is created. One XML element is identified with one \emph{metadata} object, with its attributes associated to the data members inside the object. If there is an embedded element inside the XML element, it is associated in-chain to the corresponding \emph{metadata} member. In this way, the automatic initialization of \emph{metadata} objects is achieved, without any file reading methods in the class.

\subsection{Event data processing and analysis}

\begin{figure}[tb]
  \centering
  \raisebox{-0.5\height}{\includegraphics[width=0.95\linewidth]{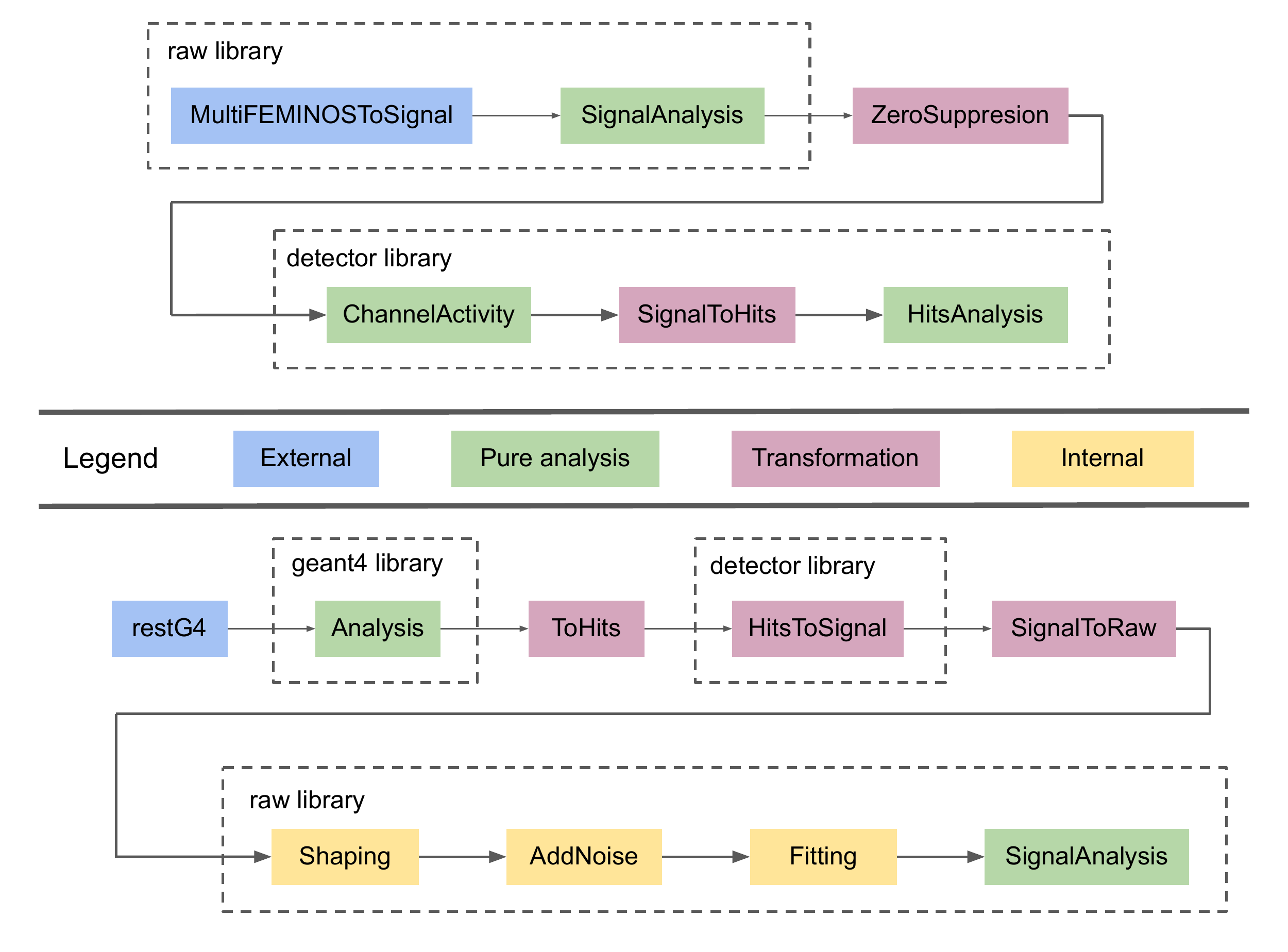}}
	\caption{A schematic showing the event data flow for two different data chain implementations in order to illustrate the different processes classification (using a color legend). On the \emph{top}, an experimental detector data processing chain reading a binary file, analyzing, and post-processing the rawdata for event reconstruction. On the \emph{bottom}, a Monte Carlo generated data processing chain, where the data are analyzed and transformed to match the data format in a raw electronics acquisition system, where the data is conditioned using \emph{shaping}, \emph{add noise} and \emph{fitting} internal processes belonging to the raw library. The schematic shows how different libraries (geant4, detector, raw, described later on section\,\ref{sec:libraries}) intervene at different stages, and how those play a role in both, Monte Carlo and experimental data.}\label{fig:dataChain}
\end{figure}

The framework allows to build an event data processing chain in a modular way by interconnecting already existing \emph{specific event processes}, or developing new ones with the potential to plug them directly to an existing processing chain. Each \emph{event process} has access to the input \emph{specific event}, the \emph{analysis tree} and any \emph{metadata} object that is accessible by the \emph{run} object. Depending on the input/output \emph{specific event} interaction inside the process, we may attempt to classify the \emph{event processes} into different groups, as illustrated in Figure\,\ref{fig:dataChain}:

\begin{itemize}
\item An \emph{external process} is a process that reads an external data source, usually at the beginning of a REST processing chain. It might be binary data generated by an acquisition system, or Monte Carlo data generated by an external simulation package. The process will be in charge of understanding the format of that external data serving to initialize a REST \emph{specific event}.

\item An \emph{internal transformation process} is a process in which the \emph{specific event} input is the same type as the \emph{specific event} output. The \emph{event} data will be transformed but not the \emph{event} type.

\item A \emph{pure analysis process} accesses the information of a \emph{specific event} type and produces observables that will be added to the \emph{analysis tree} but it will not modify the \emph{specific event} contents in any sense. A pure analysis process might serve, for example, to implement a complex physics model that uses the \emph{specific metadata} and \emph{specific event} information to elaborate some results that will be exported to the \emph{analysis tree}, or a \emph{specific metadata} object.

\item A \emph{general process} is a process that does not access the information inside the \emph{specific event} type. It will only need access to the basic \emph{event} information common to all \emph{specific events}, and/or the \emph{analysis tree}. Therefore, this process may be plugged at any point of a data processing chain without restrictions. Processes of this kind may have many different purposes, for example; to visualize online \emph{analysis tree} observables on real time, implement a \emph{summary} process to calculate averages (or any other statistical variable) from the \emph{analysis tree}, or perform a generic fitting of a variable from the \emph{analysis tree} storing the fitting results in a dedicated metadata object, among many other basic analysis tasks.

\item A \emph{transformation process} is a process that receives as input a \emph{specific event} and transforms it to a different \emph{event} type. This kind of processes will all be placed at the \emph{connectors} library, described in section\,\ref{sc:connectorslib}, in order to encapsulate all library inter-dependencies in a single entity. 
\end{itemize}

A \emph{specific event} might be transformed during the event processing and, in that transformation, relevant information might not be available anymore in the final transformed output event. The reason is that the role of the \emph{specific event} object is to provide a faithful or significant representation of the data at the state of processing inside the processing chain. At different processing stages,  the event data might be made of time signals registered at an electronics setup, or it might be in the shape of discrete energy deposits in a physical coordinate system. Therefore, the transformation from one event data representation, or \emph{specific event}, into another, means that a relevant parameter available at a particular stage, is not available anymore.

\begin{figure}[tb]
  \centering
  \raisebox{-0.5\height}{
  \begin{tabular}{c}
  \includegraphics[width=0.98\linewidth,trim=0 0 10 0, clip]{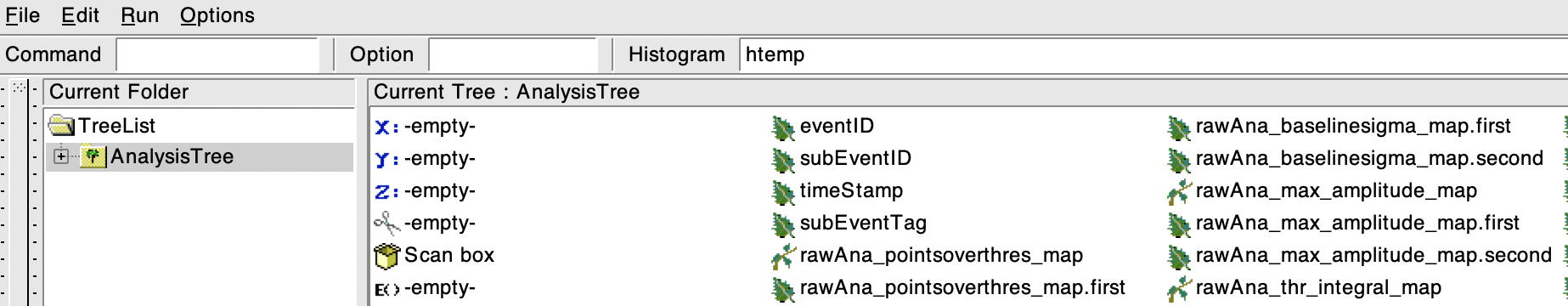} \\
  
  \\
  
  \includegraphics[width=0.65\linewidth,trim=0 98 0 0, clip]{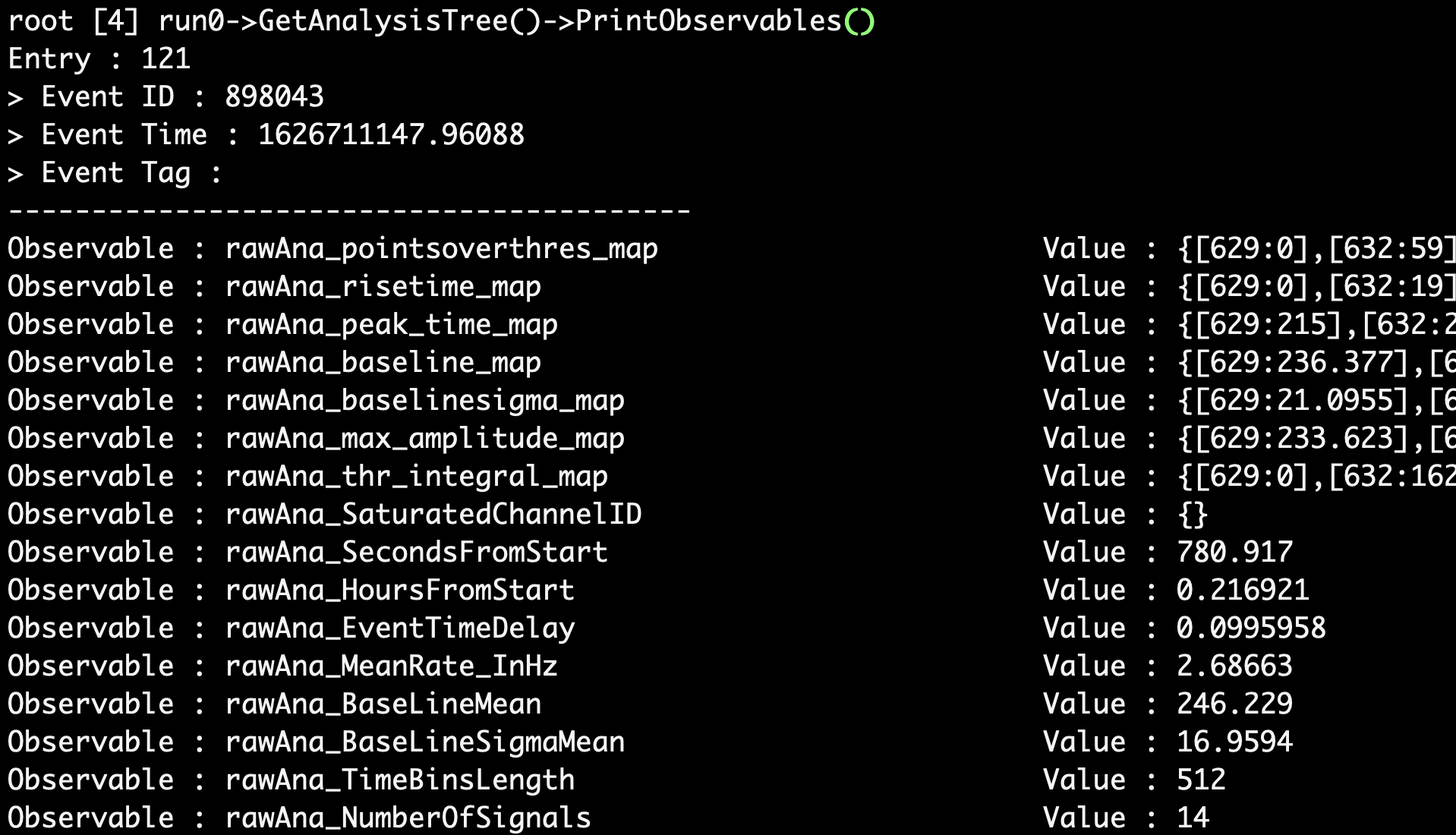} \\
  \end{tabular}
  }
	\caption{Two snapshots showing the observables registered at the analysis tree. At the \emph{top}, the observables are inspected using a ROOT browser object. At the \emph{bottom}, the analysis observables are inspected at a particular event entry using the ROOT command line interface.}\label{fig:observables}
\end{figure}

The \emph{analysis tree} comes into play as an instrument to collect all those parameters extracted or calculated from the \emph{specific event} information which will be relevant for the final analysis.  Any \emph{specific event process} in the processing chain is allowed to add new observables to the \emph{analysis tree}. Once a process adds an observable to the \emph{analysis tree}, this observable will always be available even if the \emph{event} data is transformed or the processing chain happens in several steps using different input/output REST data files. The information in the \emph{analysis tree} is always accumulative, and therefore it will contain a full summary of the observables added by each process (see Figure\,\ref{fig:observables}).

In brief, the \emph{analysis tree} provides a way for \emph{specific event processes} to export an analysis result, extracted from each event, to the framework. It must be noted that the processes have two ways to export results, an event-per-event based observable inside the \emph{analysis tree}, or a given result common to all the events in a particular \emph{run}, that will be exported in the form of a \emph{specific metadata} object.

The information extracted by a process and added to the \emph{analysis tree} might be as simple as just a registered value available at the \emph{specific event} at a given stage of the processing chain, or it might be the result of a complex calculation in the context of a physics model, including complicated input \emph{metadata} objects or parameters. Of course, even in the case of basic observables extracted directly from the \emph{specific event}, the user might be interested to know the evolution of such observable after an intermediate processing. In order to do that, it is possible to define the same \emph{specific event process} at different positions in the sequential processing chain.




The \emph{event process} class implements a method to facilitate the addition of observables to the \emph{analysis tree}. This method allows to directly create or set the value for an observable from any C++ variable\,\footnote{In principle custom data types can be implemented inside the \emph{analysis tree}, as this feature is supported in a standard ROOT tree. However, restricting ourselves to the use of standard C++ types is convenient to avoid additional dependencies and facilitate exporting the tree data outside of the framework domains.}  (supporting the most common C++ types, from base types to proper \emph{stl} containers, and either global or local variables). This method simplifies the coding of REST \emph{event processes} by avoiding users to directly interface the branch and tree ROOT objects, and at the same time it is used to encapsulate common naming conventions for observable names, or other analysis REST standard definitions.

Our framework design is completely adapted to the processing of experimental data, or Monte Carlo simulated data. The reason is that a \emph{specific event process} implementation may operate on both scenarios. The only requirement is that the experimental or simulated input \emph{event} must be given to the process in the form of a \emph{specific event} type. If both, simulation or experimental data, are conditioned to fit in a common \emph{specific event} type, it will be possible to build a processing chain that not only processes simulated data or experimental data, but that fully combines both. For example, one could integrate a process simulating the signal shaping of electronics into experimental data to assess the benefit of applying such electronics setup in our experiment. Furthermore, a proper conditioning of the generated Monte Carlo \emph{event} data will allow the evaluation of the algorithms for analysis to be used with the experimental data even before the start of the physics data taking program.

\begin{figure}[htb!]
  \centering
  \raisebox{-0.5\height}{\includegraphics[width=\linewidth]{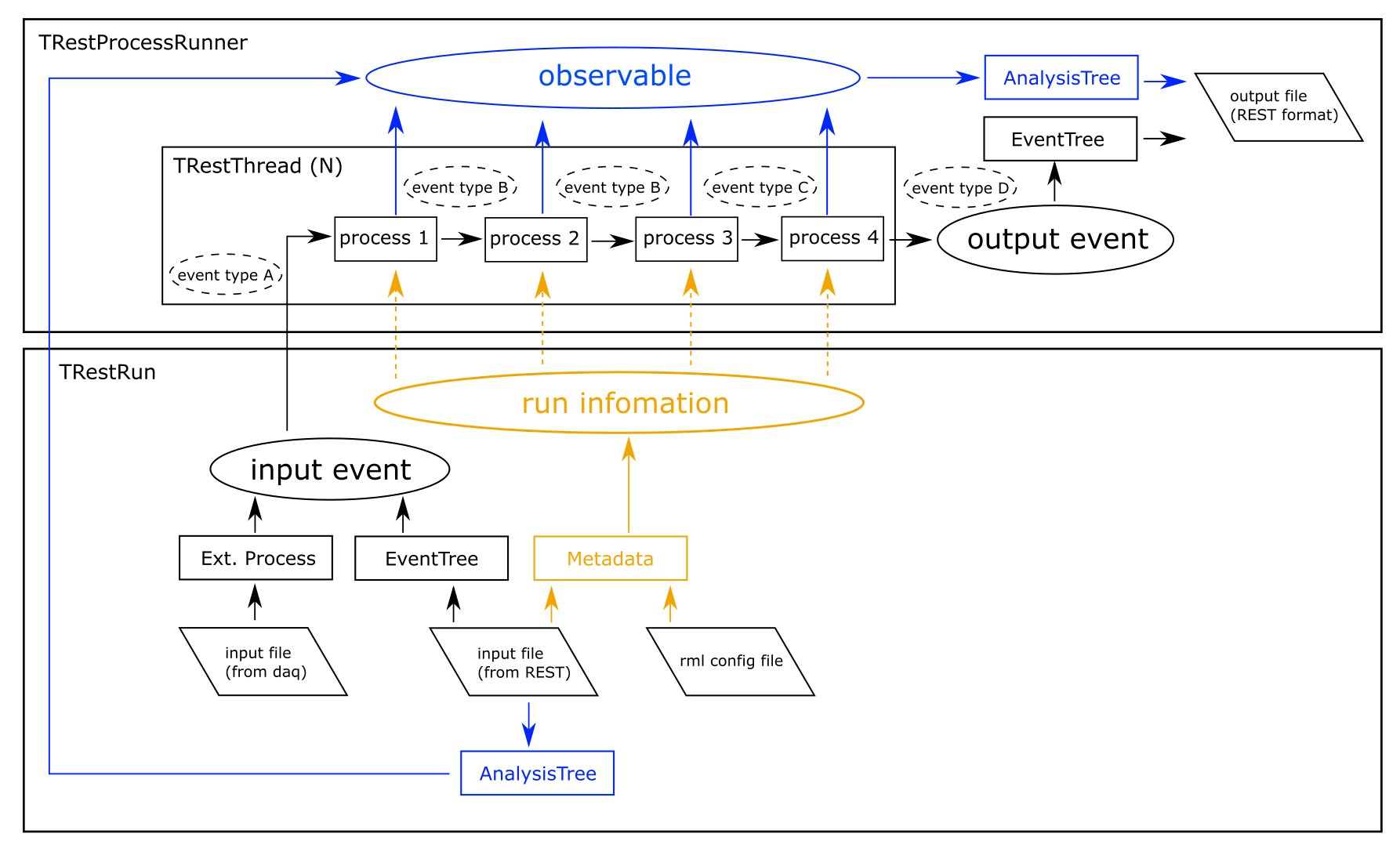}}
	\caption{A schematic diagram showing the event data flow inside a REST processing chain. The \emph{run} object is initialized, and it has access to any \emph{specific metadata} or \emph{event} data available at the input REST file, or any additional objects described through \emph{rml}. The data is then processed using the implementation inside the \emph{process runner} object. Different event types (A,B,C,D) make reference to different \emph{specific event} implementations. The resulting output REST file will contain all the \emph{metadata} information available to the chain, including any previously available, together with the transformed output \emph{specific event}, and the updated \emph{analysis tree}.  }\label{fig:processing}
\end{figure}

Event processes are executed through an efficient engine, or \emph{process runner}, with multi-thread support. The data processing chain is cloned into multiple instances and kept in different threads respectively. During execution, the input event is in turn dispatched to each thread for processing, while the output event is redirected to the global output file for writing, leading to an increase of processing speed proportional to the number of threads enabled. Figure\,\ref{fig:processing} summarizes the input/output processing logic and the different concepts already described in this section.   

\subsection{Visualization and plotting}

\begin{figure}[h]
  \centering
  \raisebox{-0.5\height}{\includegraphics[trim={1mm 1mm 1mm 1mm},clip,width=0.48\linewidth]{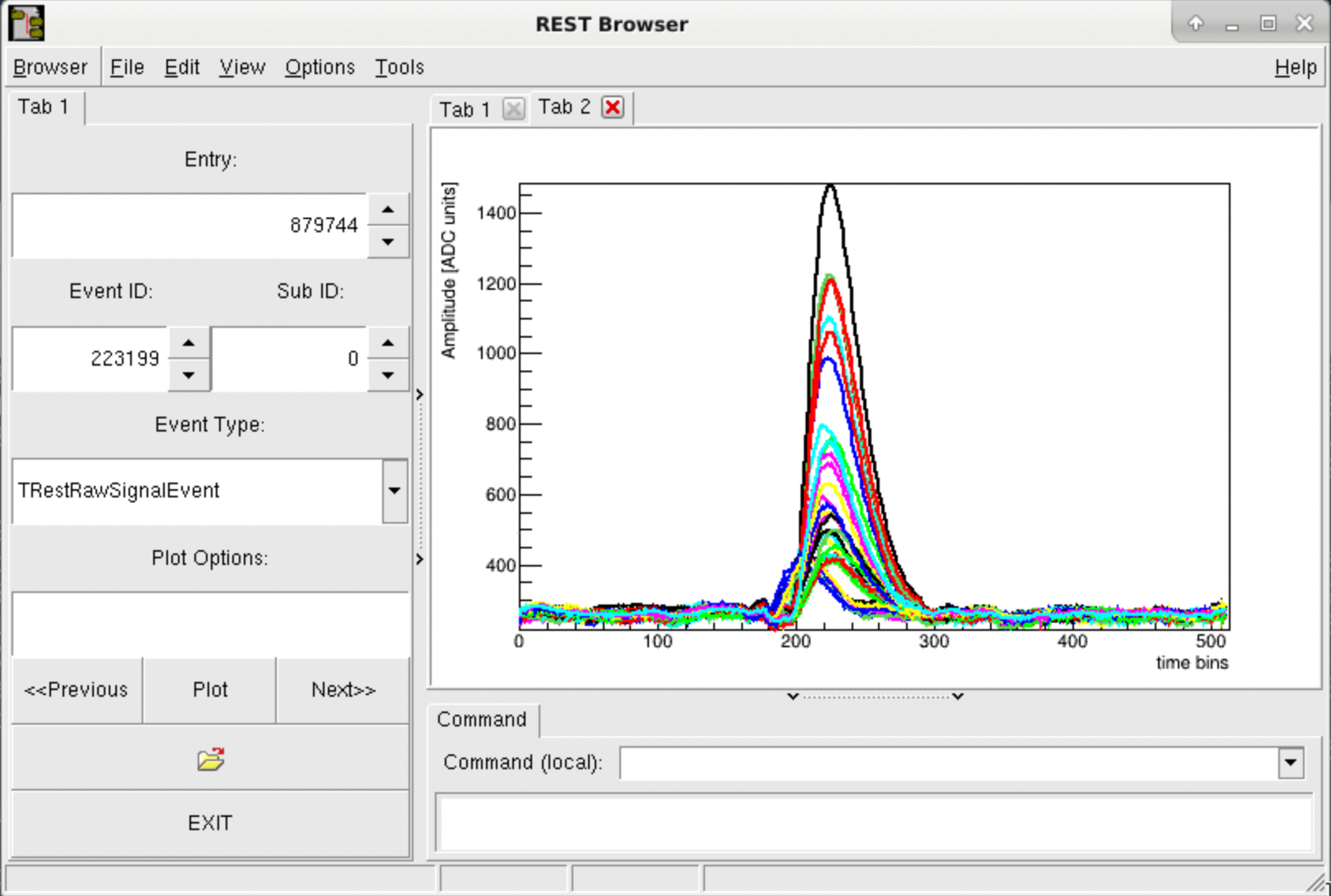}\,\,\,\,\includegraphics[trim={1mm 1mm 1mm 1mm},clip,width=0.48\linewidth]{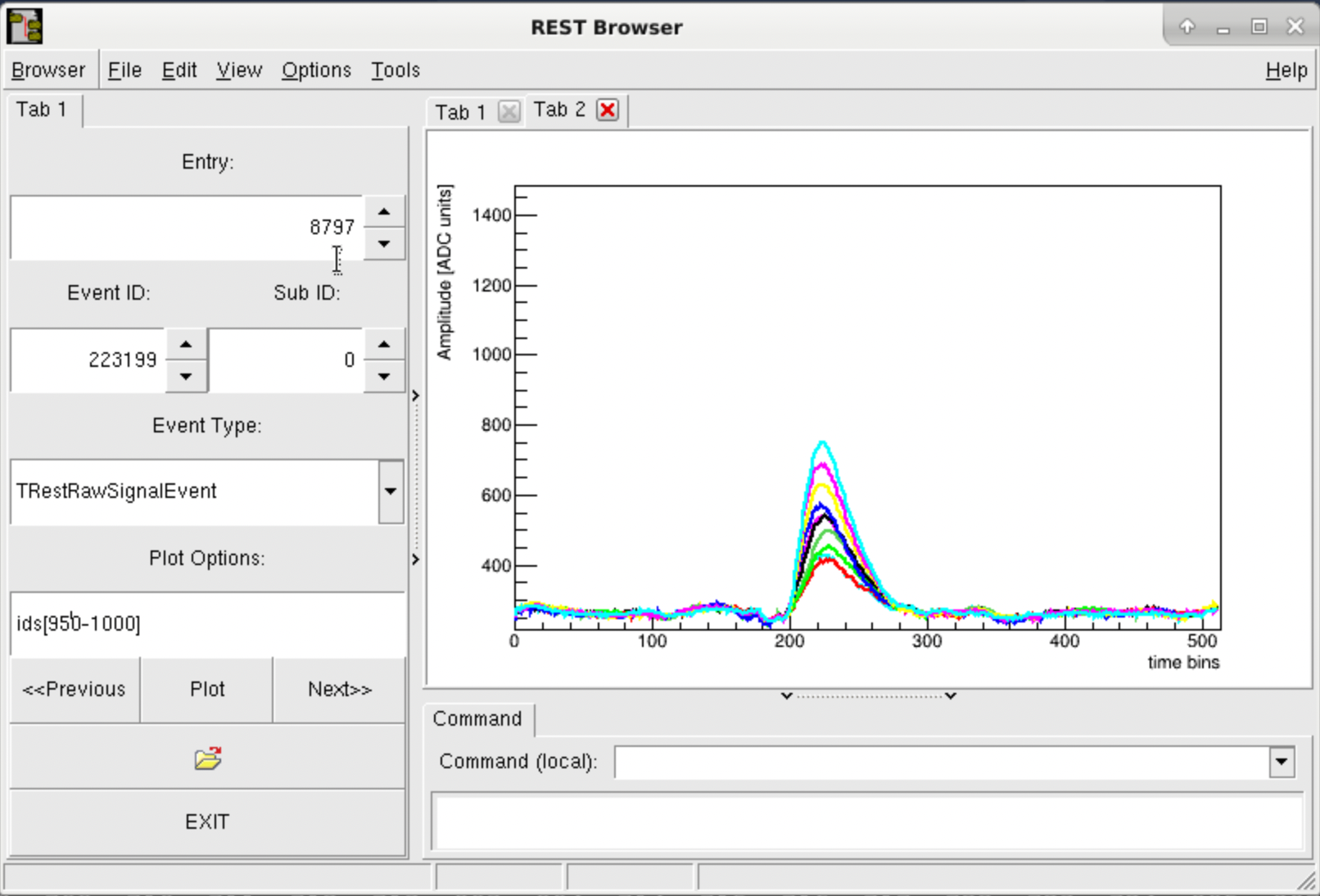}}
	\caption{Two snapshots from the REST \emph{event browser} where the control panel and the drawing pad showing an event entry for a \emph{raw signal event} is shown (see section\,\ref{sc:rawlib}). On the left figure the complete event is presented, while on the right figure the pulses have been filtered using an option passed to \emph{draw event} method, which is implemented at any \emph{specific event}.}\label{fig:eventBrowser}
\end{figure}

REST-for-Physics implements routines for event visualization and observable plotting based on ROOT drawing classes and methods. ROOT graphical interface classes are used to create basic tools, such as an \emph{event browser} with a control panel and a drawing pad (see Figure\,\ref{fig:eventBrowser}). The drawing pad itself is the target of the \emph{draw event} method implemented at each \emph{specific event}. If enabled, different output \emph{specific event} trees - from different stages in the data processing - will be stored in the same file. In that case the \emph{event browser} will be able to switch between the different event data representations.

The \emph{analysis tree} class inherits directly from the ROOT tree class, and therefore one may exploit all the resources provided by ROOT when analyzing the observables that have been added to the \emph{analysis tree} by the different \emph{specific event processes} at the processing chain: i.e. one may use a ROOT browser to explore the REST data files, and quickly draw and inspect variables from the \emph{analysis tree} (as shown previously in Figure~\ref{fig:observables}).

\begin{figure}[h]
  \centering
  \raisebox{-0.5\height}{\includegraphics[width=0.9\linewidth]{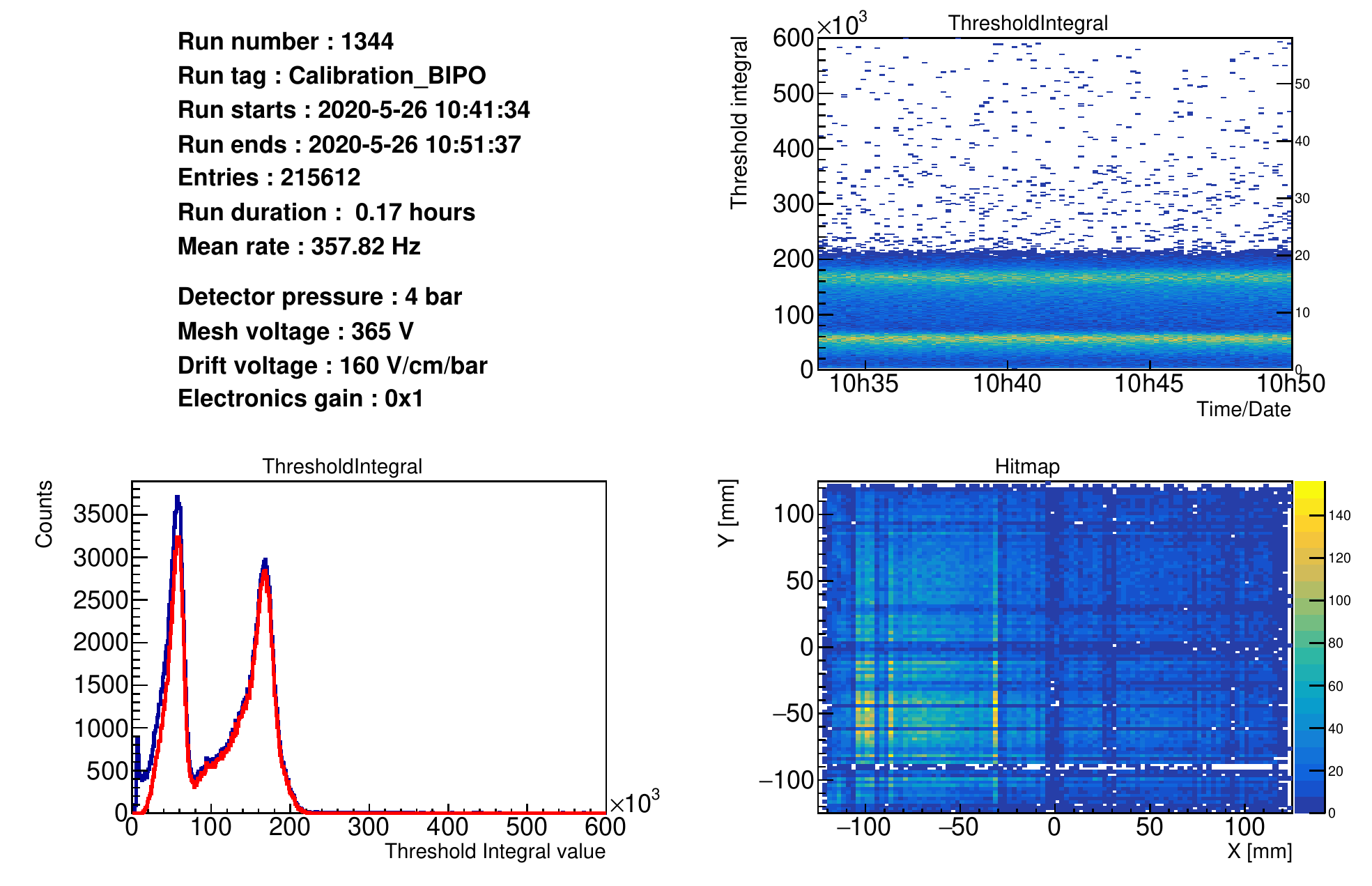}}
	\caption{A summary report produced by the quick analysis system integrated at TREX-DM using REST. The plots are generated using \emph{analysis tree} observables. A panel with \emph{run} details and other \emph{specific metadata} information (top-left), an energy spectrum (bottom-left), an energy spectra evolution along the run duration (top-right) and a distribution of the mean positions, or hitmap, where the event interactions took place (bottom-right) are shown. The \emph{ThresholdIntegral} is an observable produced by the raw library that represents the detected energy.}
	\label{fig:quickAna}
\end{figure}

Furthermore, REST implements dedicated tools for automatic and systematic plot generation, such as the \emph{analysis plot} or the \emph{metadata plot} classes. 
The \emph{analysis plot} will efficiently integrate the capability to merge thousands of files through an \emph{rml} file in which the desired plots will be assembled using the combined datasets. An \emph{analysis plot} object allows the creation of systematic plot definitions that can be used, for example, to produce quick analysis reports in a Portable Document Format (PDF), as the one produced in Figure~\ref{fig:quickAna}, or to export histogram data in any other file format supported by ROOT.
A \emph{metadata plot} object allows to read many REST generated files and draw any \emph{specific metadata} member as a function of another \emph{specific metadata} member extracted from each of the REST files provided. This enables the study of the correlation between any two metadata parameters, or the evolution of a metadata parameter as a function of the \emph{run} time, or the associated run number, for example.

\subsection{Execution and job management}
Two executables are provided at the top level of the REST-for-Physics framework and are always available to any REST user, \emph{restRoot} and \emph{restManager}: the former provides a ROOT interactive prompt with REST libraries loaded, and optionally, with all the available REST macros preloaded; \emph{restManager} manages the execution of jobs. It may launch a processing chain defined through the \emph{process runner}, execute a method defined by any REST object available to the \emph{run} object or launch a ROOT C++ macro file. 

\begin{figure}[h]
  \centering
  \raisebox{-0.5\height}{\includegraphics[width=0.9\linewidth]{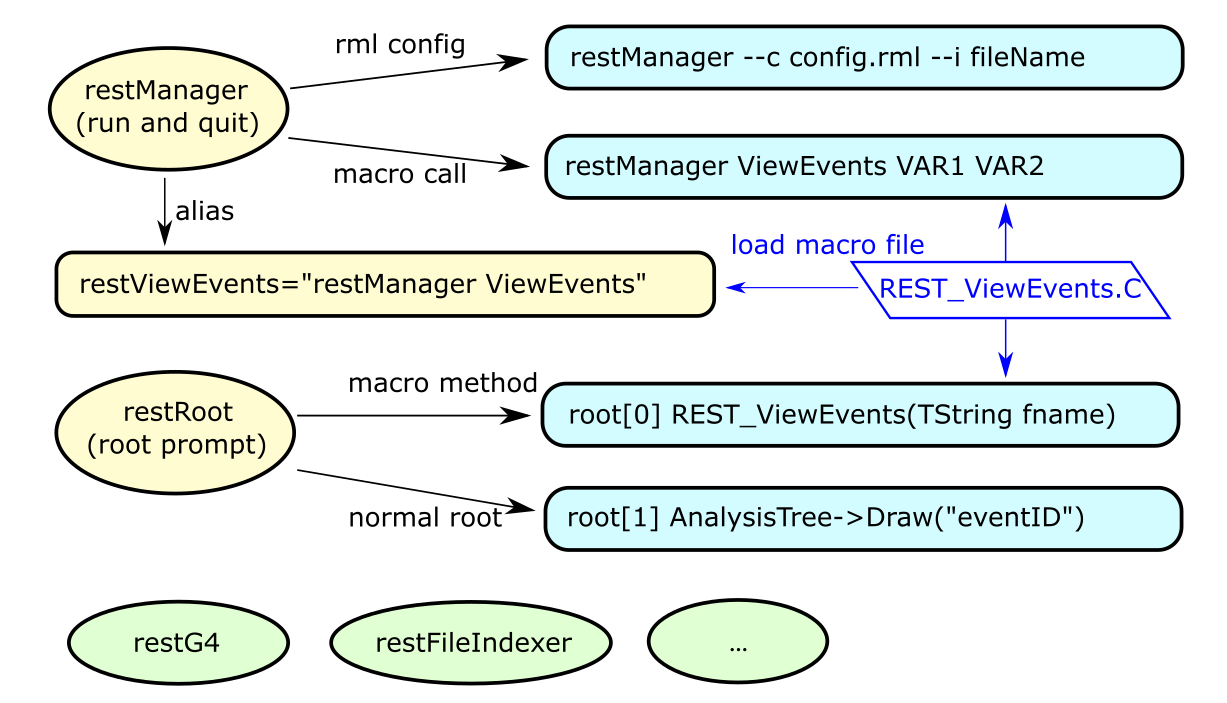}}
	\caption{REST executables running logic. \emph{restManager} and \emph{restRoot} work together to provide full access to the REST framework functionalities. Pre-defined ROOT C++ macro files are accessible through different interfaces, as it is shown through the \emph{REST\_ViewEvents.C} macro. Applications based on REST framework (green bubbles) extend the scope of the framework by providing additional functionalities, such as \emph{restG4} or \emph{restFileIndexer}.}
	\label{fig:executables}
\end{figure}

ROOT C-macros can be used to execute very specific but common tasks accessing the information inside REST data files. Official REST macros distributed with the framework may have been assigned an alias to facilitate its execution at the command line. Packages, or applications, that link to REST libraries will also provide their own executables, such as \emph{restG4} or \emph{restFileIndexer}  (see Figure~\ref{fig:executables}). \emph{restManager} allows the definition of all those actions through a configurable \emph{rml} file. The \emph{manager} class, executed through the \emph{restManager} executable, guarantees that the event data processing flow follows the standards previously described in Figure~\ref{fig:processing}.

A bash script, \emph{rest-config}, is generated at each project compilation to provide information on the configuration of a particular build and to facilitate the linking of REST with external applications. It is important to remark that once REST has been compiled with a particular version of ROOT, Geant4 or Garfield++, that compilation of REST must only be used with those versions. The shell script \emph{thisREST.sh} will be responsible to load the ROOT, Geant4, Garfield++, or any other packages required, so that they match the correct versions used to compile REST at runtime.

\subsection{Project structure, versioning and code validation}

The main framework defines the basic functions, and describes the behavior of the main elements of REST. As previously mentioned, it also serves to centralize all the REST-for-Physics components, such as packages or libraries, and eventually dedicated projects. We have adopted a \emph{git submodule}\footnote{From this point we introduce a few concepts connected with the code versioning system, \emph{git}, that are broadly available online, such as \emph{commit} or \emph{submodule}. When we refer to those alien concepts we will highlight them using the \emph{git} keyword followed by the specific concept name.} strategy to integrate those components in a modular way inside the main framework repository. This scheme allows to independently monitor the development activity at each of those components, to isolate technical issues, and to focus on their functionality. Each component evolves independently with its own version or tracking system. A particular state of the code at each of those components is fixed at the main framework through a \emph{git commit} hash, or a unique number. When that happens, the corresponding \emph{git commit} becomes the official component version of REST.

The framework repository fully centralizes the versioning system of REST, understood as the state of the code at a given period of time, including the state of the official \emph{git submodules} attached to it. Any REST \emph{metadata} object written to disk using the ROOT I/O scheme will be stamped with metadata values (e.g. the REST release number, latest commit hash, release date, etc) that ensure that the data written to disk has been processed with a given version, or state of the code. In order to certify that, two of those metadata members will be initialized at the code compilation time. The first metadata member will guarantee the source code was built from a clean, unmodified state with respect to the \emph{git remote} repository, and the second metadata member will certify that the corresponding framework code state is associated with an official \emph{git tag} release, where each \emph{git tag} generated at the main framework repository will automatically produce a code release referenced and citable at the Zenodo system\,\cite{javier_galan_2021_5092550}.

On top of that versioning strategy, it is important to mention that REST properly implements the ROOT schema evolution and ensures backwards compatibility for objects that have suffered changes in their data members.



\begin{figure}[h]
  \centering
  \raisebox{-0.5\height}{\includegraphics[width=0.95\linewidth]{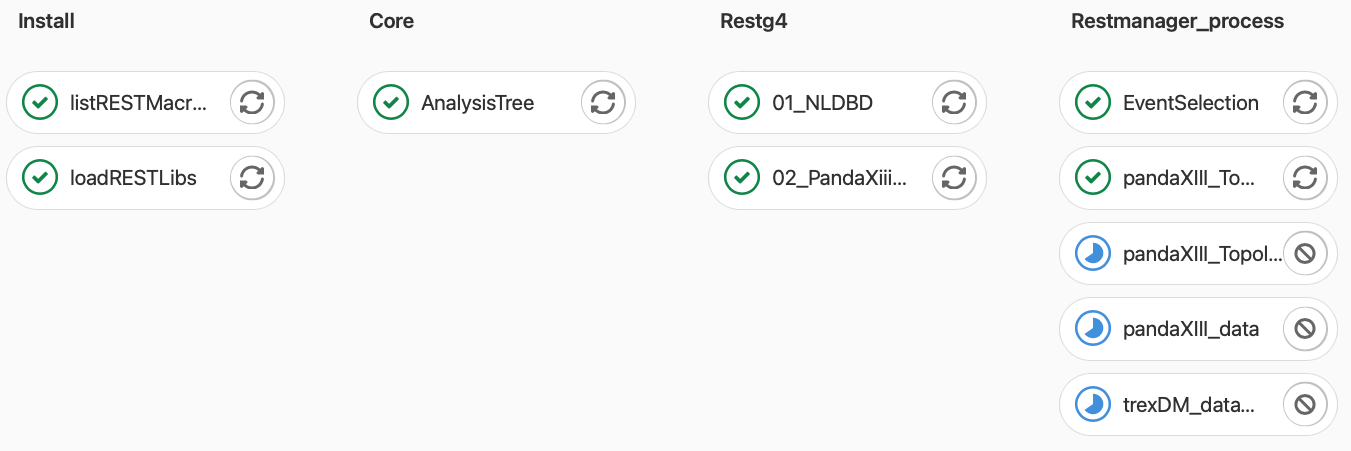}}
	\caption{A snapshot from a validation pipeline at \emph{gitlab.cern.ch} running different tests triggered by an update to the code at the main framework repository. Different validation stages are shown, from the most basic tests on the left, including compilation and installation to complex data chain processing tests on the right.}\label{fig:pipelines}
\end{figure}

To ensure the code quality and stability with time, each repository integrates a validation pipeline where basic tests on the code are performed: some examples are code formatting and style validation, testing the proper libraries integration and building of executable programs or, even more important, testing basic results from complex data processing chains (see Figure\,\ref{fig:pipelines}). Each modification to the code, or \emph{git commit}, will be verified by running those validation pipelines. If a modification to the code produces an unexpected value on a consolidated data processing routine, the contributor will be notified, and changes will only become official after peer reviewing the code. This fact is extremely relevant to guarantee that the algorithms keep producing the expected results, or in the undesired case of a bug code identification, promptly identify the affected routines after its correction. Moreover, validation pipelines might serve as running examples to show the integration or use of a specific tool or element operating inside the framework.

\section{REST-for-Physics libraries}

\label{sec:libraries}

The main framework contains common tools required for centralized data access, visualization, and basic analysis routines, including generic REST-for-Physics \emph{metadata} classes and \emph{processes} that do not require \emph{event} specialization, i.e. they only need to access information at the \emph{analysis tree} level. More specialized routines, requiring a dedicated \emph{event} data type, such as time signal processing or detector event reconstruction, are organized into libraries; all classes belonging to the library keep a closer relation and therefore enhanced connectivity.

A library is usually associated only with one or two \emph{specific event} types, increasing the connectivity between different \emph{specific event processes} inside the same library. In this way, any combination of processes belonging to a particular library can be connected inside a data processing chain within its library domain. A dedicated library, the \emph{connectors} library, hosts those \emph{specific event processes} or \emph{specific metadata} objects that need to interconnect different libraries, keeping all inter-library dependencies bound together into a single entity and allowing each library to be fully operational in stand-alone mode.

A class belonging to a particular library will have its library name as a prefix at the class name. Therefore, the \emph{TRest} naming convention is extended in the case of the libraries to \emph{TRestLibName}, enabling the prompt identification of the library an object belongs to\footnote{In this context, we will continue highlighting the words that make reference to C++ objects using that pattern, such as \emph{TRestDetectorReadout} being written as \emph{detector readout}, or even omitting the library keyword, writing, for example, \emph{TRestDetectorGas} simply as \emph{gas}.}.

Even though new libraries might be added in the future to the framework, this section briefly describes those fundamental libraries that gave REST-for-Physics enough functionality and versatility to be used in different aspects of rare event searches experiments.

\subsection{The detector library}

The detector library\,\cite{REST_Detector_Git} has been designed to be used for event reconstruction inside a Time Projection Chamber (TPC) filled with a gaseous medium\footnote{The currrent version of REST-for-Physics has only been exploited with gaseous TPCs. However, a liquid TPC or even other detector technologies will probably share common detection elements, like the generic \emph{detector readout} implementation, or several detector physics processes. }. This library contains metadata class definitions that allow to describe the detector configuration: these can be \emph{drift volume} description, the \emph{detector readout} topology, the particular \emph{gas} properties (extracted using the Magboltz interface implemented by Garfield++) or others. It also integrates processes implementing routines for event reconstruction from real detector data and/or emulation of different physical response effects, e.g. including \emph{electron diffusion}, or artificially introducing the detector energy resolution by means of a \emph{smearing} process.

\begin{figure}[htb!]
  \centering
  \raisebox{-0.5\height}{
  \begin{tabular}{ccc}
  \includegraphics[width=0.3\textwidth]{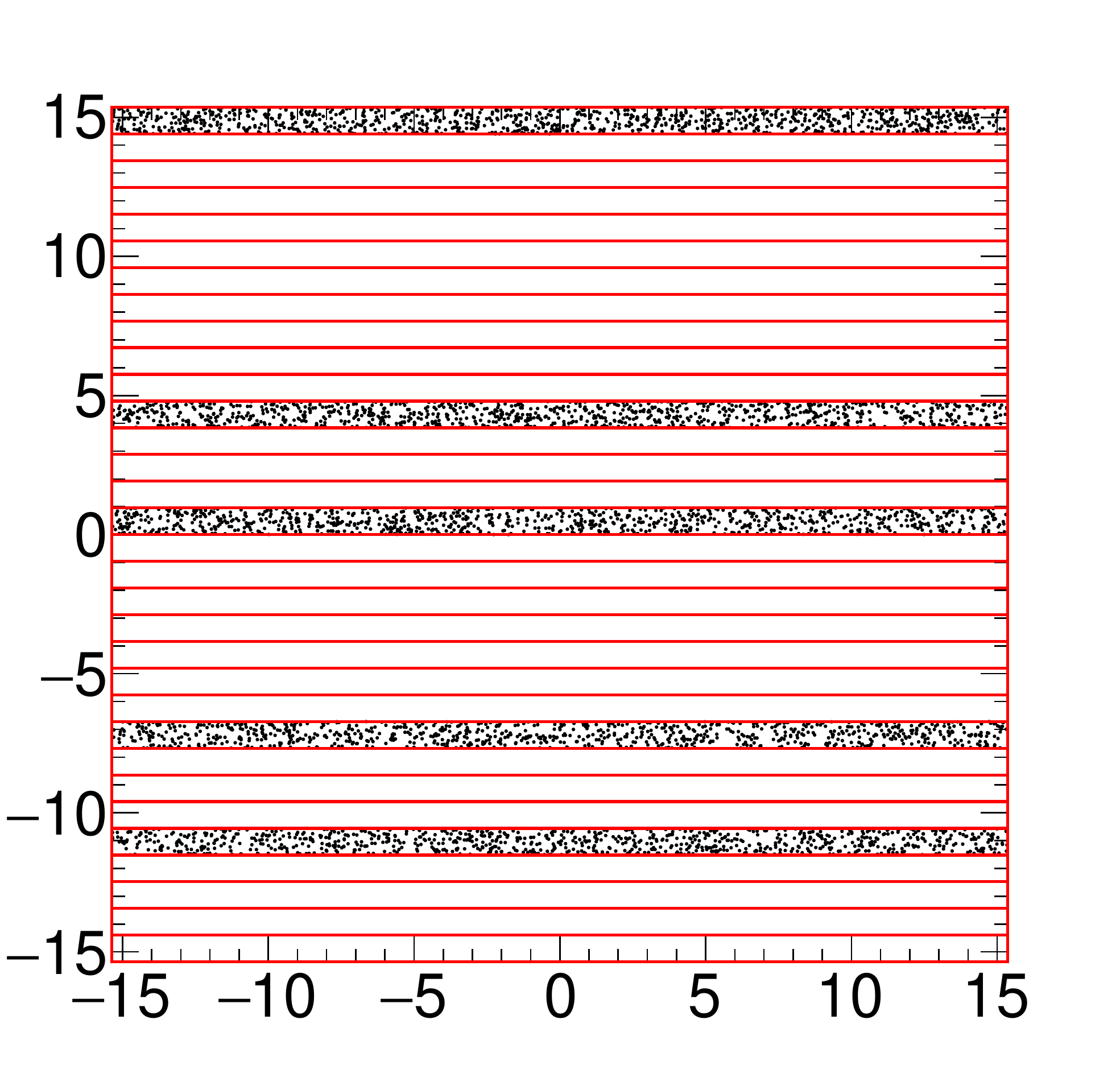} & \includegraphics[width=0.3\textwidth]{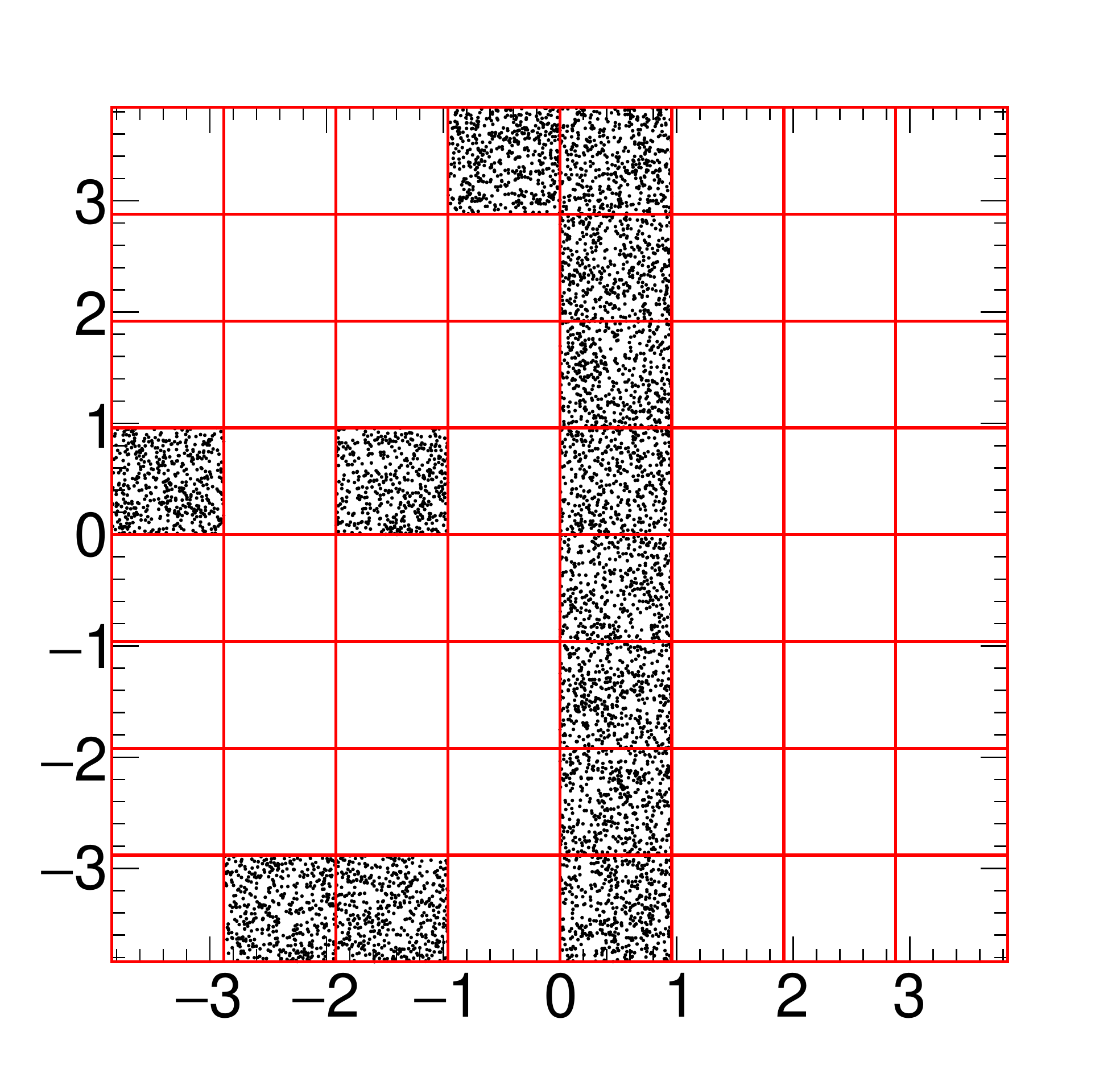} & \includegraphics[width=0.3\textwidth]{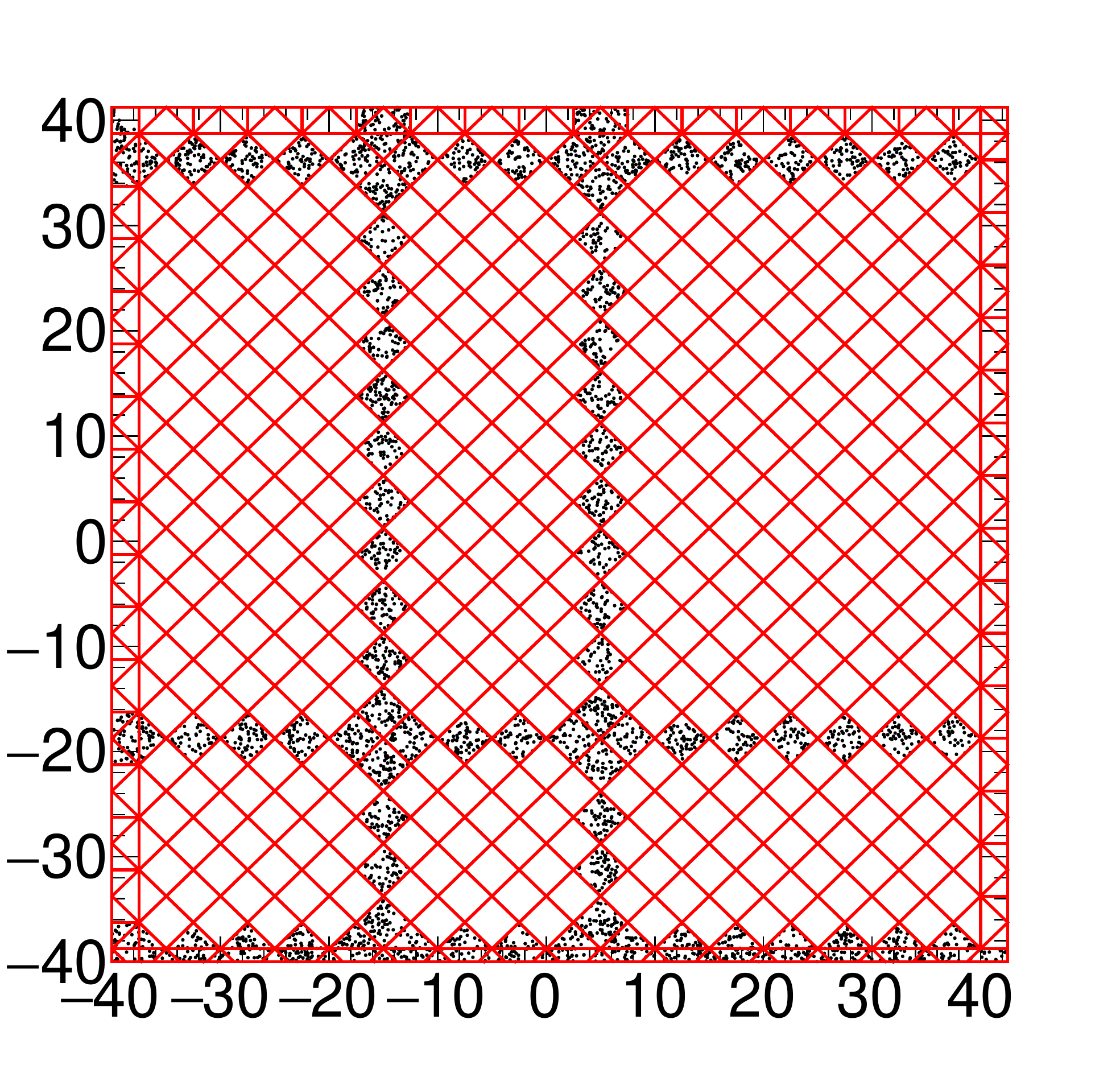} \\
  (a) & (b) & (c) \\
  \end{tabular}}
	\caption{Basic readout topologies that can be found at the \emph{basic-readouts} repository\,\cite{REST_Readouts_Git}. (a)~A stripped \emph{readout channel} layout. (b) A pixel layout. (c) A more complex layout where each \emph{channel} is composed of a few interconnected \emph{readout pixels} that create a stripped pattern. The red lines represent the boundaries of the \emph{readout pixels}, while the black dots are produced by launching a randomly spatial distribution and drawing only those points that fall inside dummy user-enabled \emph{channels}, serving for \emph{readout} design validation.}\label{fig:readouts}
\end{figure}

The \emph{readout} construction (see Figure\,\ref{fig:readouts}) is a crucial element of the detector library. This element permits the definition of an arbitrary number of \emph{readout planes}, containing an arbitrary number of \emph{readout modules}, composed of physical \emph{readout channels} that identify unambiguously with the acquisition channels of an electronics setup. The \emph{readout channels} are at the same time built with \emph{readout pixels}, the most basic element of a \emph{detector readout}. Such scheme allows to create any arbitrary and complex topology, with the capability to efficiently translate - back and forward - physical coordinates and electronic channels for readouts containing a few millions of pixels.

As any other library, the detector library provides an event type to encapsulate the detector data. Currently, and for convenience, it is the only library that defines two event types. The \emph{detector hits event} type, and the \emph{detector signal event} type. The \emph{hits event} defines a physical quantity, the energy deposits at the detector physical volume, using a 3-dimensional spatial coordinate representation. The \emph{signal event} describes the energy deposits as a function of the arrival time to the \emph{readout plane} associated to each detector electronics channel. The \emph{readout} implementation works as a dictionary between those two types; it is used to translate one event type into another by projecting the energy deposits into the \emph{readout channels}, or by recovering back the physical coordinate description from the readout channels information.

This library plays a central role in the characterization of the detector data and thus naturally includes connections to REST libraries related to raw electronics data processing (section\,\ref{sc:rawlib}), particle physics Monte Carlo event processing (section\,\ref{sc:geant4lib}) or physical track identification and pattern recognition routines (section\,\ref{sc:tracklib}). The processes responsible for such library inter-connectivity are hosted on an independent library, the connectors library (see section\,\ref{sc:connectorslib}).

\subsection{The raw library}\label{sc:rawlib}

The raw library\,\cite{REST_Raw_Git} implements a \emph{raw signal event} type that is suited to describe the time evolution of physical quantities that have been acquired with a fixed sampling rate. Inside this event type one may find an arbitrary number of \emph{raw signals} that, in the case of TPC technology, are identified with the induced currents in the electronic channels. Each \emph{raw signal} inside the event definition contains usually the same number of samples, a value which is fixed during the \emph{raw signal} initialization. The data depth of the physical quantity described inside the \emph{raw signal} is 16-bits precision, which is enough to fit the typical values of electronic acquisition systems.

\begin{figure}[htb!]
  \centering
  \raisebox{-0.5\height}{
  \begin{tabular}{cc}
  \includegraphics[width=0.48\textwidth]{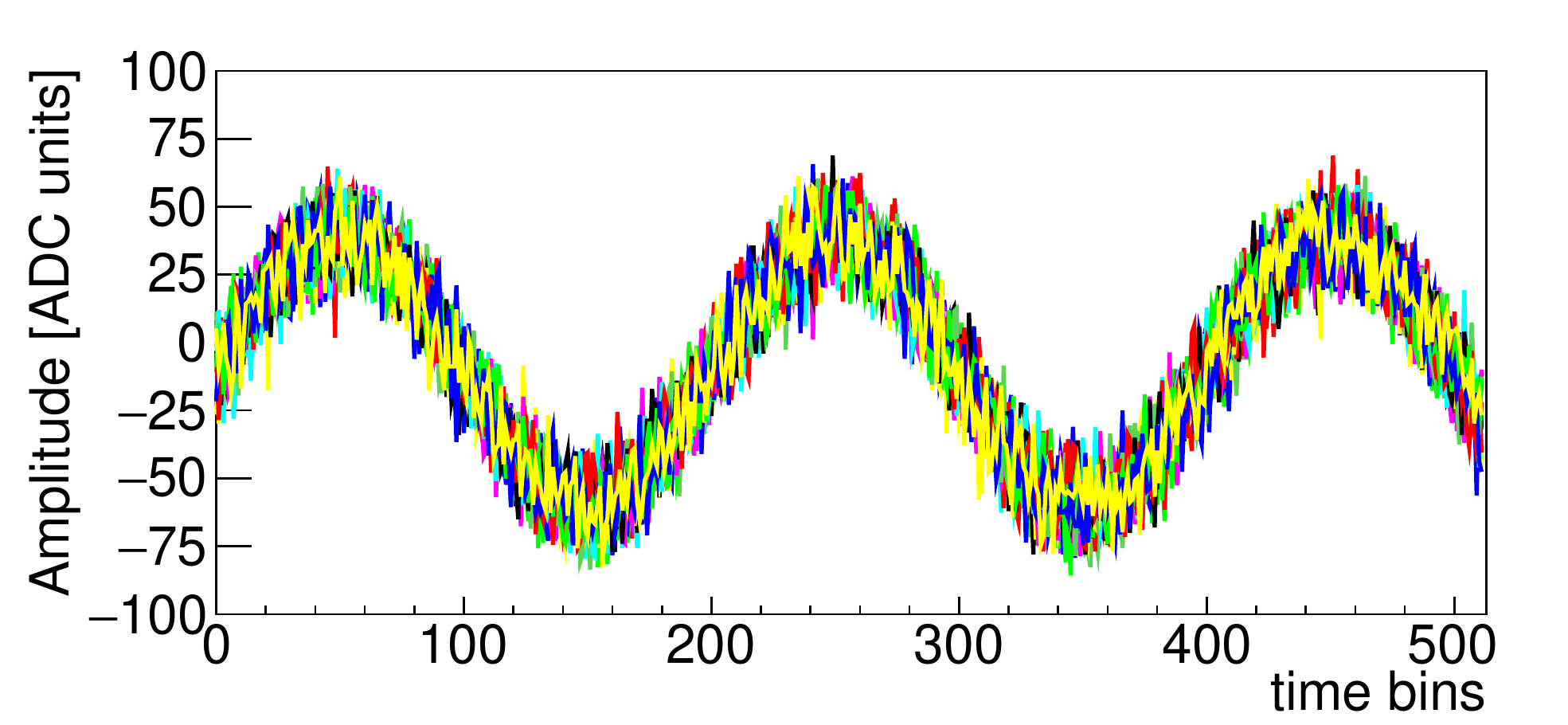} &
  \includegraphics[width=0.48\textwidth]{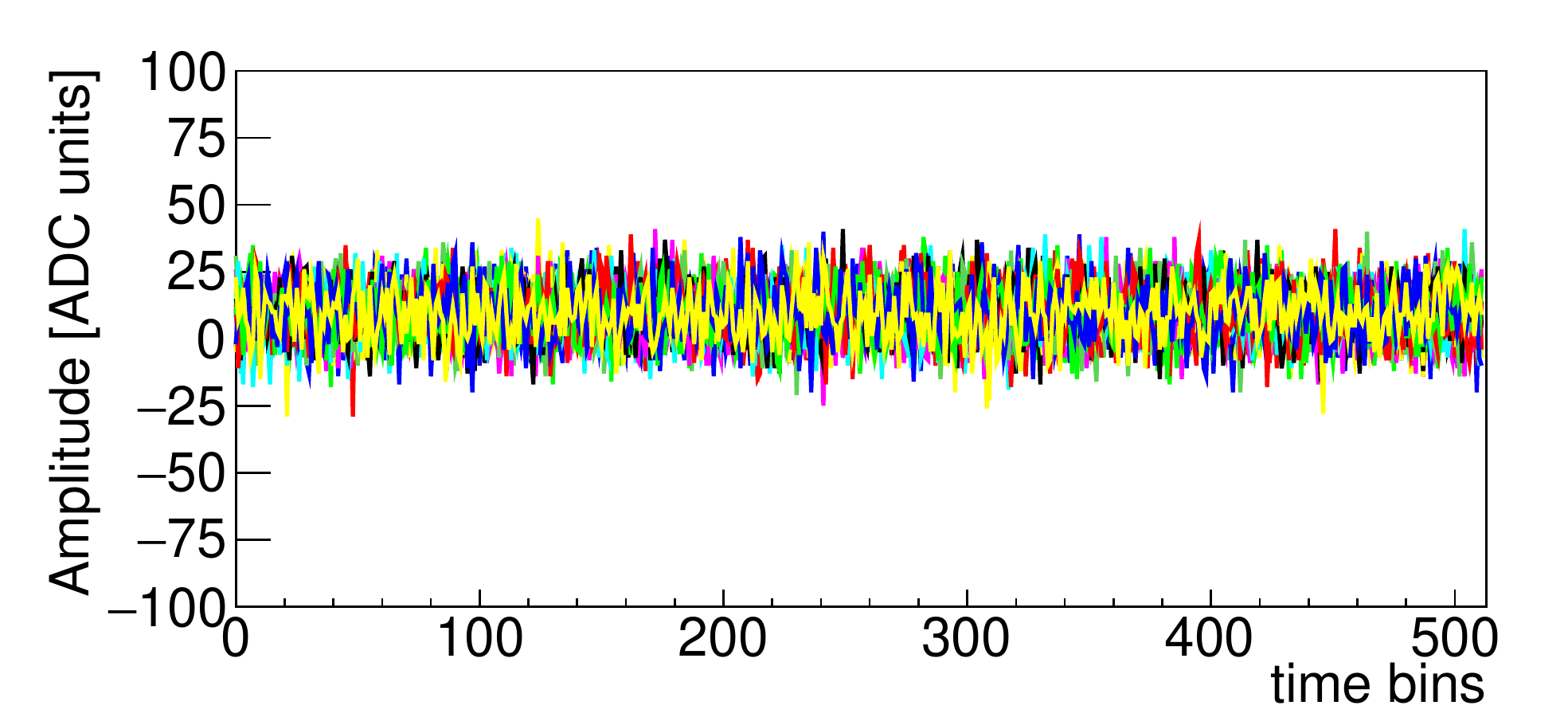} \\
  (a) & (b) \\
  
  \includegraphics[width=0.48\textwidth]{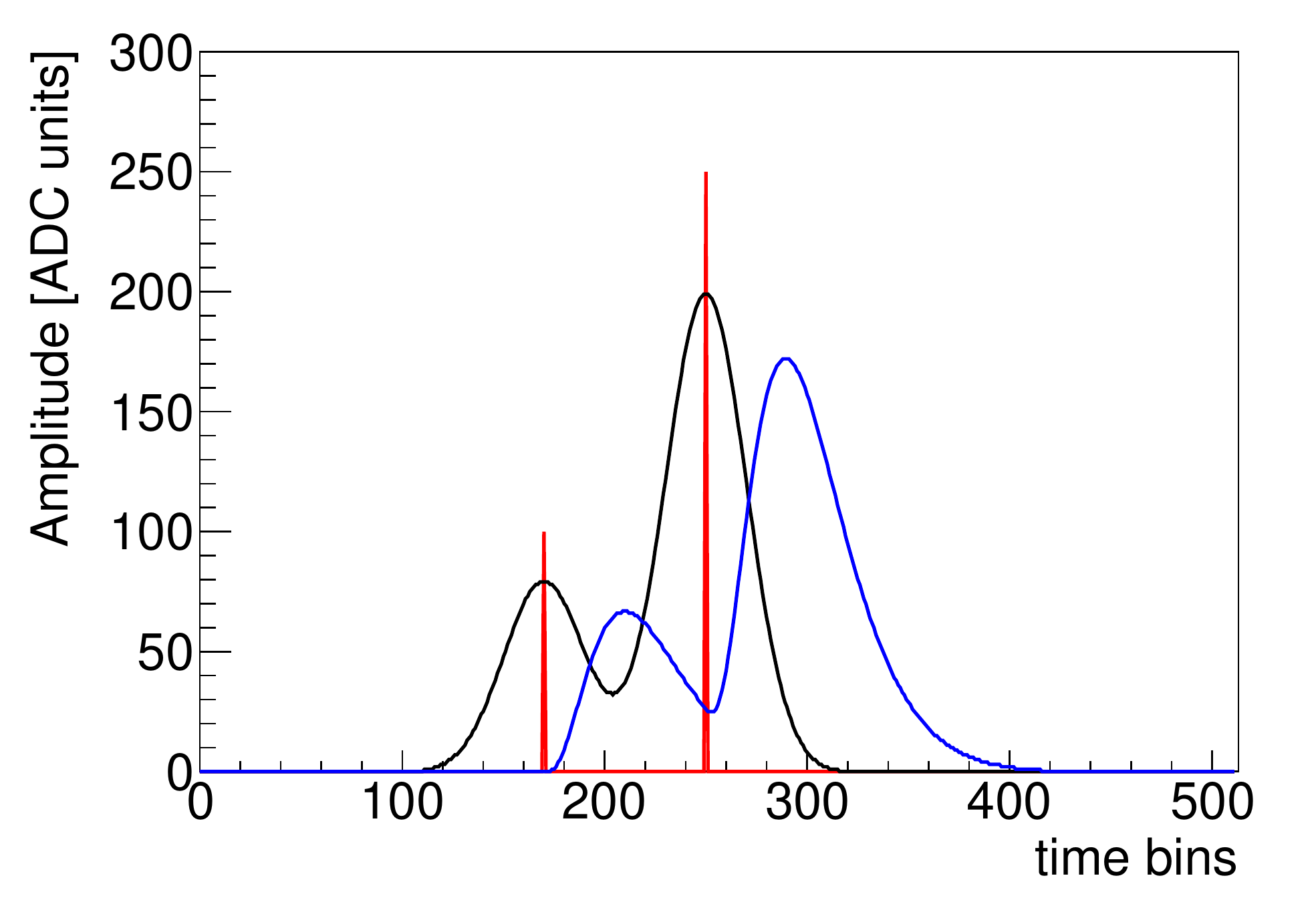} &
  \includegraphics[width=0.48\textwidth]{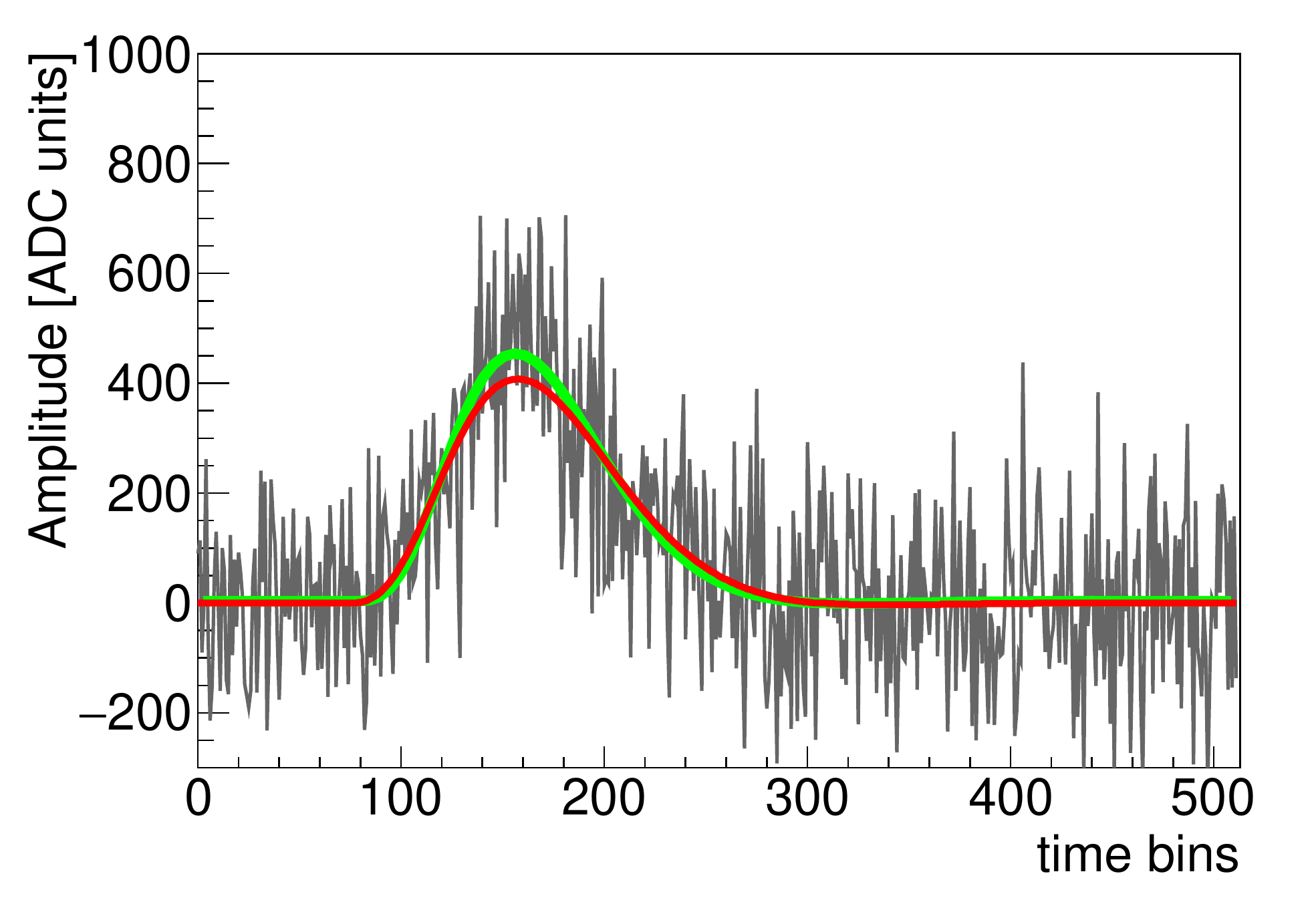} \\
    (c) & (d) \\
  \end{tabular}
  }
	\caption{(a) An artificially-generated noise-\emph{raw signal event} with a common sinusoidal pattern. (b) The result after applying the \emph{common noise reduction} process to the event shown in (a), where only randomly added noise remains. (c) An idealized \emph{raw signal} composed of two point-like deposits (in red) is conditioned by the \emph{shaping} process using a Gaussian (in black) and an AGET electronics response (in blue) convolutions. (d) An artificially-generated noise-\emph{raw signal} (in black), together with the original pulse used to generate it (in red), and the recovered \emph{raw signal} after applying the \emph{fitting} process (in green).}\label{fig:rawlib}
\end{figure}

This library includes processes related to signal conditioning, such as signal shaping, de-convolution, pulse fitting, de-noising, Fast Fourier Transform operations, common noise reduction and other signal manipulation routines in the time domain (see Figure\,\ref{fig:rawlib}). 

In addition, the raw library includes processes, belonging to the external process type, that allow to import into the framework the binary data generated by different electronics acquisition cards used in our field, such as AGET\,\cite{6154095} and AFTER\,\cite{Baron:2009sdx} chips, or DREAM\,\cite{Acker2020} electronics, among others.

\subsection{The geant4 library}\label{sc:geant4lib}

The geant4\footnote{We will use the lowercase version of the \emph{geant4} word when we refer to our own REST-for-Physics code implementation, while we will use the upper-case version, Geant4, to refer to the official CERN software package~\cite{Agostinelli:2002hh}. As a reminder, highlighted words provide a connection with the code objects, as \emph{geant4 event} being linked to the object \emph{TRestGeant4Event}.}  library\,\cite{REST_Geant4_Git} defines a \emph{geant4 event} type that registers the energy deposits, or hits, resulting from a Geant4 simulation. A Geant4 simulation performs the physics particle tracking including the interaction probability with the materials defined for a given detector geometry. The energy deposits are similar to those found at a \emph{detector hits event}, although the \emph{geant4 event} hits contain additional information, like the physical interaction process, the geometrical volume where the interaction took place or the remaining available kinetic energy of the particle that produced the energy deposit. The energy deposits are encapsulated into \emph{geant4 tracks} that describe properties common to a particular group of hits, such as the particle name producing the energy deposits, the position where the particle was originated, the track and parent ids, and in general, any relevant information directly extracted from the tracks produced by the Geant4 simulation package.

It is important to mention that this library is not directly linked to the official Geant4 libraries. Its purpose is to store the event information generated by a Geant4 simulation, but once a simulation package has registered the information inside the \emph{geant4 event} data holder, the connection to Geant4 libraries is not required anymore. Therefore, a user would be able to access a Monte Carlo database of previously Geant4-generated files in REST format without the need to perform a system Geant4 installation.

\begin{figure}[htb!]
  \centering
  \raisebox{-0.5\height}{\includegraphics[width=0.4\textwidth]{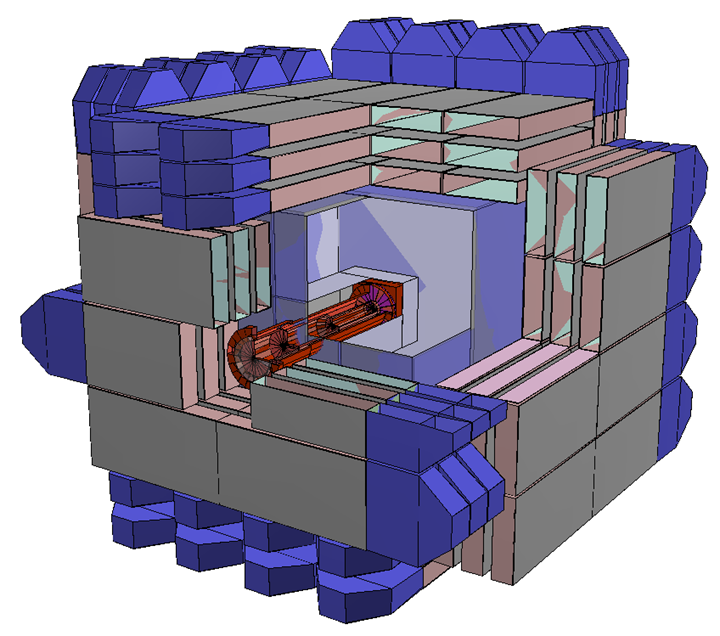}\hspace{1 cm}\includegraphics[width=0.35\textwidth]{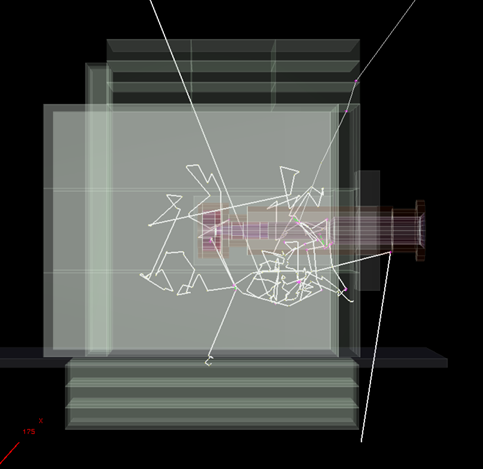}}
	\caption{\emph{Left}, a visualization of the GDML geometry for the Baby-IAXO detector\,\cite{BabyIAXO:2020mzw}. \emph{Right}, a simulated cosmic neutron event in the same geometry visualized using the ROOT TEve viewer libraries.}\label{fig:geant4lib}
\end{figure}

Inside the REST-for-Physics ecosystem we have developed an independent package, \emph{restG4}\,\cite{REST_restG4_Git}, which is a particular Geant4 code implementation taking advantage of the \emph{geant4 event} type and all the definitions available at the library to describe the simulation conditions. For example, the \emph{geant4 metadata} class defines the number of primaries to be generated, together with their energy and angular distributions, or the generator type, in order to determine how the primaries will be launched or initialized. There are many other options that allow to produce datasets in different experimental conditions and apply specific storage instructions. The library implements another relevant metadata object, the \emph{geant4 physics list}, in which the particle physics processes to be considered in the simulation can be customized. \emph{restG4} will register those metadata structures and the \emph{geant4 event} tree, together with a \emph{run} metadata object complying with the REST data format conventions so that the resulting data are ready to be further processed with this or other libraries available in REST. A simulation with \emph{restG4} requires as input the description of those three objects, the \emph{run}, the \emph{geant4 metadata} and the \emph{geant4 physics list}, through an \emph{rml} file, and a description of the geometry through a GDML\,\cite{Chytracek:2006be} file (see Figure\,\ref{fig:geant4lib}).

Once a first Monte Carlo dataset has been generated using \emph{restG4} it can be processed using the existing routines in this library. These routines, or processes, can be used to extract the Monte Carlo truth at an early processing stage. One example is the \emph{blob analysis} process, aiming to extract the real electron track-ends in a $0\nu\beta\beta$ event: another is the \emph{neutron tagging} process allowing to produce elaborated observables (e.g. the mean position of energy deposits found at a particular volume in the geometry) to perform a detailed analysis of the interaction of neutrons with an active cosmic veto system. Therefore, some processes at this library introduce sophisticated physics models producing results that will be exported to the \emph{analysis tree} in the form of observables, to be accessed at a later stage of data treatment. The main idea, or philosophy, is that \emph{restG4} is simply used to generate a first dataset, while the geant4 library will be used to introduce models that need to know about the nature of the particles or the interactions that produced the energy deposits inside the detector geometry. Once all the relevant information has been extracted and placed in the form of observables in the \emph{analysis tree} it can be migrated to other REST libraries (see section\,\ref{sc:connectorslib}) in order to include a detailed detector response, condition the data to mimic raw detector data, and perform the same data processing and analysis applied to real experimental data.

\subsection{The track library}\label{sc:tracklib}

The track library\,\cite{REST_Track_Git} implements a \emph{track event} type that defines inheritance relations between a set of \emph{tracks} stored inside the event. A \emph{track} itself contains a group of hits (or cluster) that define a discrete energy distribution in a 3-dimensional coordinate space. In order to produce or initialize a first \emph{track event}, a process in the connectors library (section\,\ref{sc:connectorslib}) makes use of the \emph{detector hits event} as input to identify groups of hits, or energy deposits, that have a proximity relation, in order to create \emph{tracks}. It is important to remark that the \emph{track event} is an abstract object\footnote{Not to be confused with an abstract C++ class (it would have been highlighted otherwise). We want to emphasize that it is an object that does not have a strict or fixed scope.} that allows to define groups of hits, clusters, with an inheritance relation, i.e. one may develop \emph{track} levels by generating new daughter \emph{tracks} from the original ones. This could be exploited in different contexts: it could serve to describe isolated clusters (or group of hits) in a single physical volume, or it could serve to describe correlated \emph{tracks} from independent physical volumes by creating a new \emph{track} that incorporates all those mother tracks into one.

\begin{figure}[htb!]
\centering
    \raisebox{-0.5\height}{\includegraphics[width=0.33\textwidth]{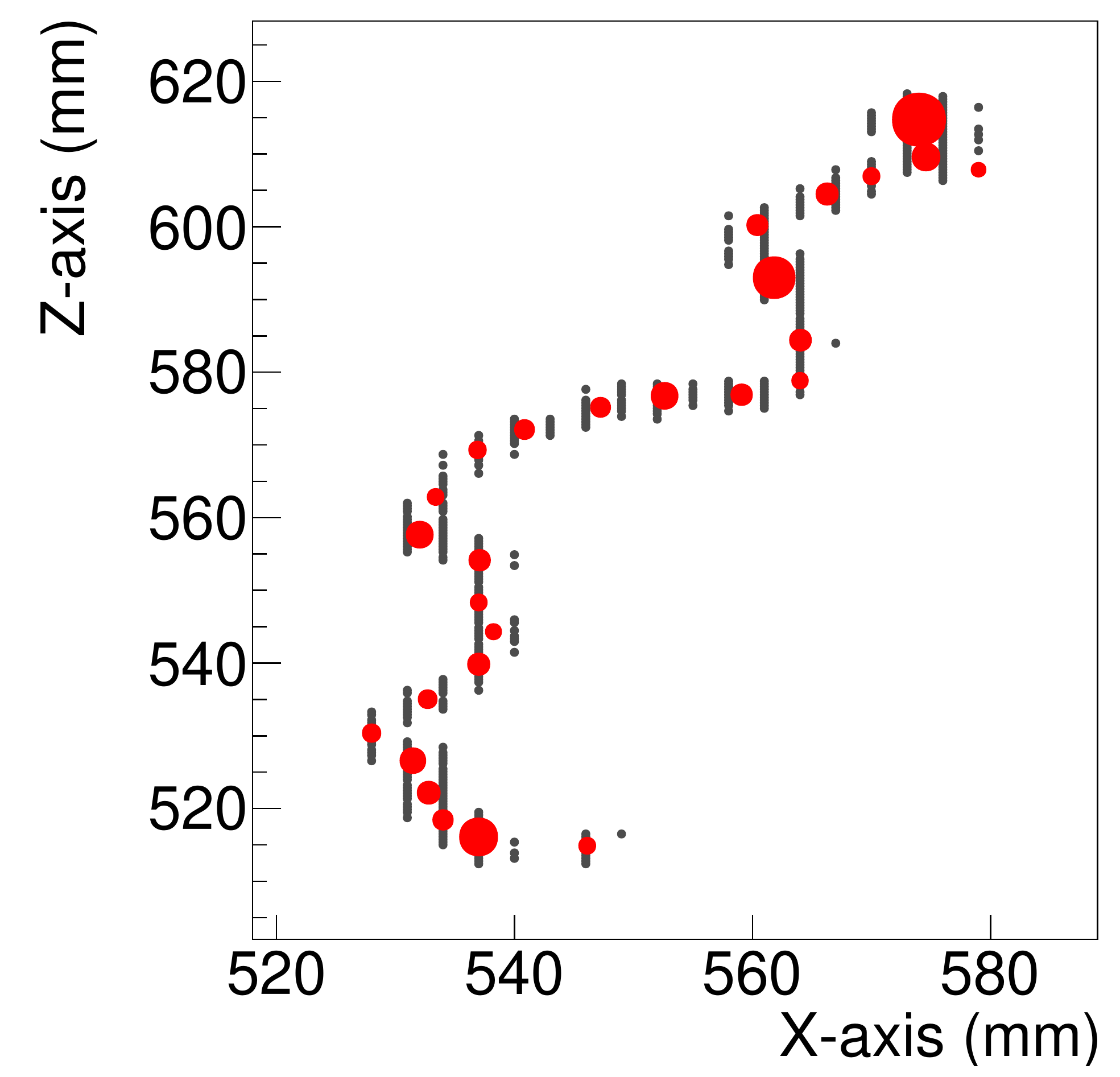}
    \includegraphics[width=0.33\textwidth]{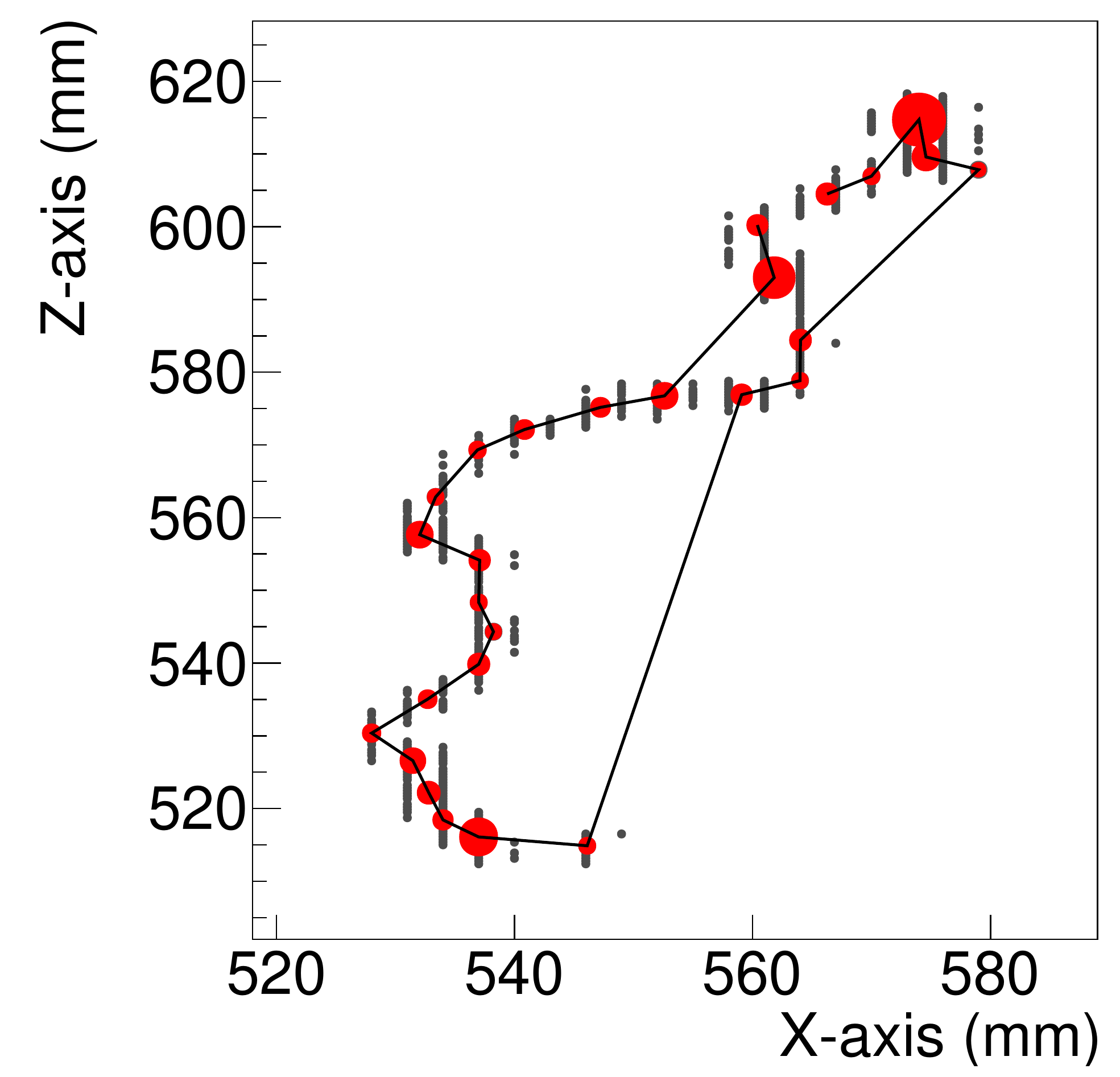}
    \includegraphics[width=0.33\textwidth]{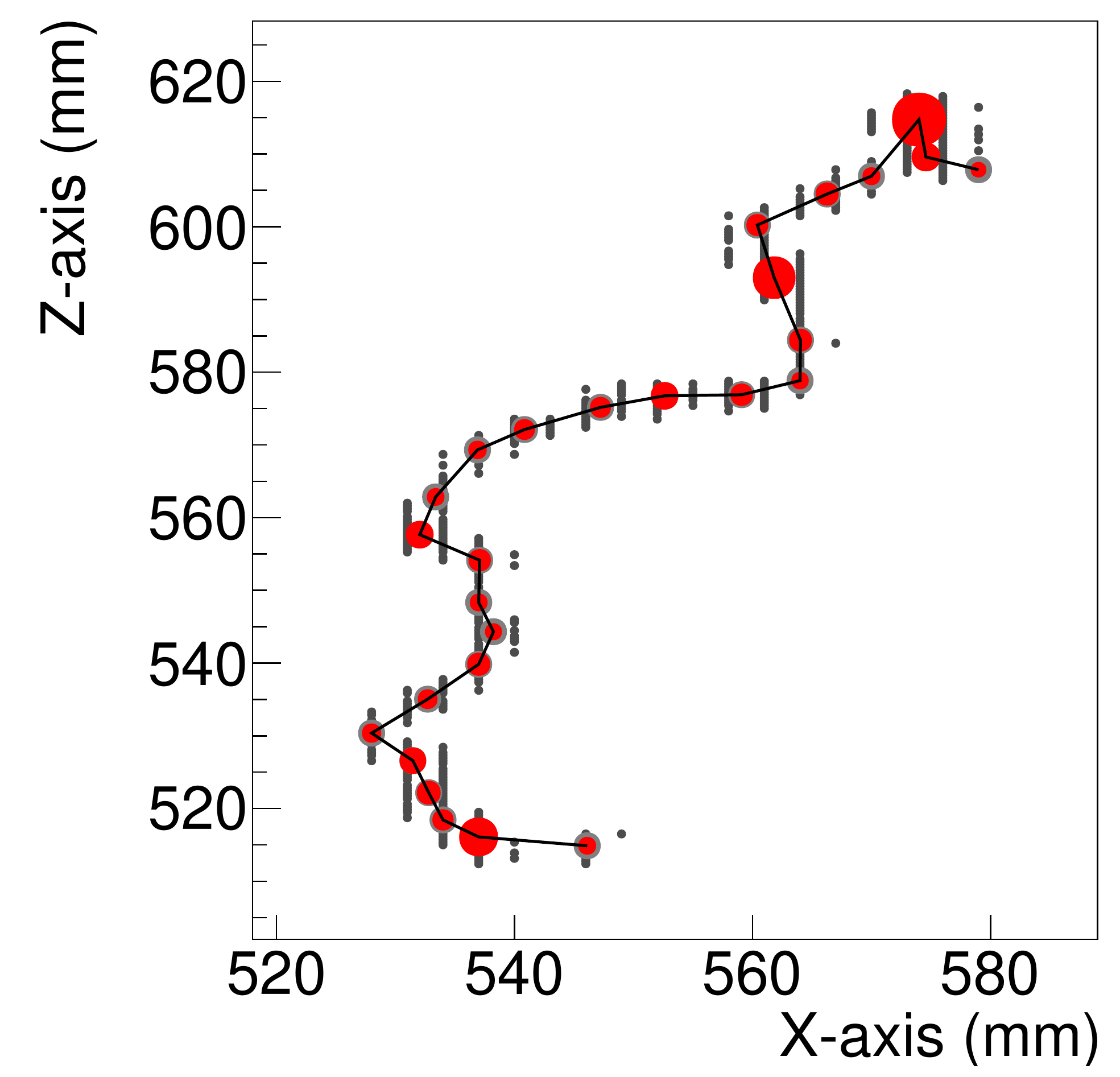}}
    \caption{A \emph{track event} representation of a simulated $0\nu\beta\beta$ decay after the treatment with different \emph{track} processes used for physical track identification. \emph{Left}, an image of the hit reduction produced by the \emph{track reduction} process. The red circles represent the final position of reduced hits, whose size is weighted with their energy value. The small grey circles on the background represent the hits of the \emph{parent track} used as input. \emph{Middle}, a polyline is added to this representation to visualize the hits inter-connectivity after the \emph{track path minimization} process. If path minimization works on the whole, it produces at times obviously unphysical connections, as our example illustrates. \emph{Right}, the unphysical connections are corrected using a \emph{track reconnection} process. Figure extracted from reference\,\cite{Galan:2019ake}.}
    \label{fig:tracklib}
\end{figure}

This library contains, on one hand, graph theory algorithms helping to identify and reconstruct physical tracks by finding the shortest path that interconnects energy deposits within a \emph{track}, and on the other, processes that allow to extract topological information from a \emph{track event}. Since graph theory algorithms are computationally expensive when dealing with a large number of nodes, a \emph{reduction} process can be used to decrease the effective number of \emph{hits}, so that Traveling Sales Problem (TSP) algorithms can be applied in an acceptable computation time\,\cite{Applegate:2007:TSP:1374811,concorde}. TSP methods help to find a reasonable solution for the physical track identification, although further \emph{reconnection} algorithms may be needed to improve the result (see Figure\,\ref{fig:tracklib}). An important application of these algorithms is the identification of neutrinoless double beta decays, as it has been shown in the context of the PandaX-III experiment\,\cite{Galan:2019ake}.

\subsection{The connectors library}\label{sc:connectorslib}

\begin{figure}[htb!]
  \centering
  \raisebox{-0.5\height}{\includegraphics[trim=0 130 0 130, clip, width=0.75\linewidth]{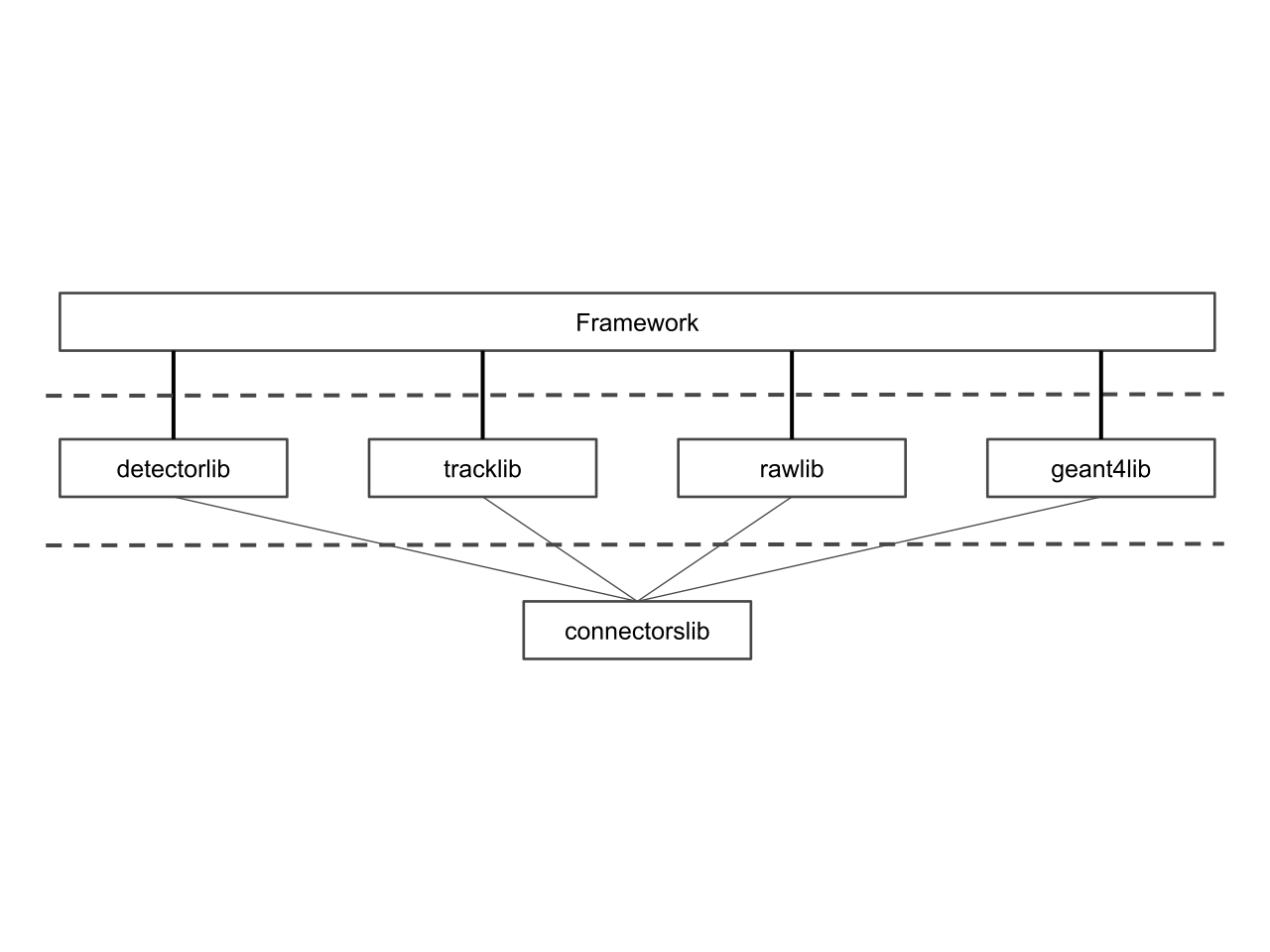}}
	\caption{REST-for-Physics libraries hierarchy and connectivity to the framework. The connectors library depends on the other fundamental libraries, providing class definitions that help inter-library communication. On the other hand, fundamental libraries with a direct connection to the framework are capable to operate in a stand-alone mode, without any other REST-for-Physics libraries requirements.}\label{fig:connectorslib}
\end{figure}

The connectors library\,\cite{REST_Connectors_Git} contains class definitions that need to combine the features from classes residing in different REST-for-Physics libraries. This includes processes that transform the \emph{event} type from one library specific \emph{event} type into another library \emph{event} type, or it includes complex \emph{metadata} object descriptions that require combining specific metadata descriptions from different libraries. The main mission of this library is to keep inter-library dependencies isolated or encapsulated in a single entity. In this way the fundamental libraries described in previous sections will be operative in a stand-alone mode philosophy (see Figure\,\ref{fig:connectorslib}). The REST-for-Physics building system will compile only those connectors library classes related to libraries that were marked for compilation: in the extraordinary case that only a single library was marked, then the connectors library will not be compiled at all. This library differentiation helps the coherent development of independent libraries. Using this design any library may be enabled or disabled at will, avoiding unnecessary dependencies on dedicated systems.

The main functionality of this library is to allow moving from one fundamental library domain into another, e.g. transforming a \emph{raw signal event} into a \emph{detector signal event} by using data reduction techniques, or grouping hits inside a \emph{detector hits event} to produce a \emph{track event}. However, the connectors library must not be understood as a simple \emph{event} data type transformation, since the \emph{specific event} data usually requires sophisticated routines that include the detector physics involved for the event reconstruction, data reduction inside signal processing algorithms or graph theory for the clustering of hits. This library will play a crucial role to define how different library domains inter-connect.



\section{Summary}

In this work we have given a broad overview of the REST-for-Physics framework and components. Our aim was to provide the reader with a general idea of the philosophy, structure and organization of the software project. And, without entering into great detail, provide an overview of the present use and functionality of REST-for-Physics.

The REST-for-Physics framework and libraries are a natural extension of ROOT, since the most basic elements inherit directly from TObject. The ROOT I/O serialization is exploited to manage the data storage while focusing on the development of physical processes that provide to REST its functionality. The motivation for this choice is the experience acquired with the ROOT framework, and the benefit of using the analysis tools it provides. ROOT was born already more than 25\,years ago and it is still strongly supported and actively maintained by the CERN community which counts with thousands of users. ROOT is exhaustively used in particle physics today, and its continuity in the long term seems to be guaranteed by CERN.

The REST-for-Physics framework fully exploits the schema evolution from ROOT in order to minimize the impact on data member changes in \emph{specific event} or \emph{metadata} objects, thus making files written with REST to be backward- and forward-compatible. One of the key aspects of the REST-for-Physics code, crucial for the storage and processing of experimental data, is its versioning strategy that it was carefully described in this paper. Such versioning strategy provides a unique relation between the code and the registered data, ensuring data and code traceability, leading to reproducible results.

One of the main motivations of the development of REST-for-Physics is to  collect and centralize the software efforts and progress on detector physics for the construction of low-background detection technologies. As such, REST-for-Physics aims to serve as a platform to support future contributions in the field, consolidating common processing routines on event reconstruction, signal conditioning or pattern recognition. REST has been widely tested using gaseous TPCs, although its routines share many aspects with other detector technologies: some of the routines could be directly exploited by other technologies, while others would require minor changes to be useful for other detection setups.

\section*{Acknowledgements}
We acknowledge support from the the European Research Council (ERC) under the European Union’s Horizon 2020 research and innovation programme, grant agreement ERC-2017-AdG788781 (IAXO+), and from the Spanish Agencia Estatal de Investigaci\'on under grant FPA2016-76978-C3-1-P. The IRFU group acknowledges support from the Agence Nationale de la Recherche (France) ANR-19-CE31-0024.

\bibliography{paper}

\end{document}